\documentclass[aps,prd,tightenlines,showpacs,amsmath,amssymb]{revtex4}%
\usepackage{graphicx}
\usepackage{epsfig}
\usepackage{graphics,color}
\usepackage{amssymb}
\usepackage{amsbsy}
\newcommand{\be}{\begin{equation}}
\newcommand{\ee}{\end{equation}}
\newcommand{\bse}{\begin{subequations}}
\newcommand{\ese}{\end{subequations}}
\newcommand{\bea}{\begin{eqnarray}}
\newcommand{\eea}{\end{eqnarray}}

\newcommand{\half}{\frac{1}{2}}

\newcommand{\bean}{\begin{eqnarray*}}
\newcommand{\eean}{\end{eqnarray*}}

\begin{document}
\preprint{}

\title{Thermodynamic properties  and bulk viscosity near phase transition\\
 in the $Z(2)$ and $O(4)$ models} 

\author{Bao-Chun Li$^{1,2}$\footnote{
libc@ihep.ac.cn}, Mei Huang$^{2,3}$ \footnote{huangm@ihep.ac.cn}}
\affiliation{$^{1}$ Department of Physics, Shanxi
University,Taiyuan Shanxi, China\\
$^{2}$  Institute of High Energy Physics, Chinese Academy of
Sciences, Beijing, China\\
$^{3}$ Theoretical Physics Center for Science Facilities, Chinese Academy of Sciences, Beijing, China }

\begin{abstract}
We investigate the thermodynamic properties including equation of state, 
the trace anomaly, the sound velocity and the 
specific heat, as well as transport properties like bulk viscosity in
the $Z(2)$ and $O(4)$ models in the Hartree approximation of 
Cornwall-Jackiw-Tomboulis (CJT) formalism. We study these properties 
in different cases, e.g. first order phase transition, second 
order phase transition, crossover and the case without phase transition,
and discuss the correlation between the bulk viscosity and the thermodynamic 
properties of the system. We find that the bulk viscosity over entropy 
density ratio exhibits an upward cusp at the second order phase transition, 
and a sharp peak at the 1st order phase transition. However, this 
peak becomes smooth or disappears in the case of crossover. This 
indicates that at RHIC, where there is no real phase transition and 
the system experiences a crossover, the bulk viscosity over entropy 
density might be small, and it will not affect too much on hadronization.  
We also suggest that the bulk viscosity over entropy density ratio is 
a better quantity than the shear viscosity over entropy density ratio 
to locate the critical endpoint. 
\end{abstract}


\maketitle

\section{Introduction}

Studying Quantum chromodynamics (QCD) phase transition and properties of 
hot/dense quark matter at high temperature and baryon density has been the main 
target of heavy ion collision experiments at the Relativistic Heavy Ion collider 
(RHIC), the forthcoming Large Hadron Collider (LHC) and FAIR at GSI. 

At small baryon chemical potential $\mu$, for QCD with two massless quarks, 
the spontaneously broken chiral symmetry is restored at finite temperature, 
and it is shown from lattice QCD \cite{CEP-lattice} and effective QCD models 
\cite{CEP-models} that this phase transition is of second order and belongs 
to the universality class of $O(4)$ spin model in three dimensions 
\cite{Pisarski-Wilczek}. For real QCD with two quarks of small mass, the 
second order phase transition becomes a smooth crossover at finite temperature. 
At finite baryon chemical potential, there are still no reliable results from 
lattice QCD due to the severe fermion sign problem. However QCD effective models 
\cite{CEP-models} suggest that the chiral phase transition at finite $\mu$ 
is of first order. It is expected that there 
exists a critical end point (CEP) in the $T-\mu$ QCD phase diagram. The CEP is 
defined as the end point of the first order phase transition, and belongs to 
the $Z(2)$ Ising universality class \cite{Uni-CEP}. The precise location of the CEP
is still unknown. In the future plan, RHIC is going 
to lower the energy and trying to locate the CEP. The signature of CEP has 
been suggested in Refs. \cite{Sig-CEP}. Recently, the authors of 
Ref. \cite{etas-CEP} suggested using the shear viscosity over entropy 
density ratio $\eta/s$ to locate the CEP .

The ratio of shear viscosity over entropy density $\eta/s$ has attracted 
a lot of interests.
It was expected that deconfined quark matter formed at high temperature should 
behave like a gas of weakly interacting quark-gluon plasma (wQGP).  It is
now believed that the system created at RHIC is a strongly coupled quark-gluon 
plasma (sQGP) and behaves like a nearly "perfect" fluid \cite{RHIC-EXP,RHIC-THEO}.
One crucial quantity is the shear viscosity over entropy density $\eta/s$, which
is required to be very small and close to the the lower bound
$\eta/s=1/4\pi$ \cite{bound} to fit the elliptic flow at RHIC from hydrodynamic 
simulation  \cite{Hydro}. Lattice QCD calculation confirmed that $\eta/s$ for the 
purely gluonic plasma is rather small and in the range of $0.1-0.2$ \cite{LAT-etas}. 
The perturbative QCD calculation gives a large shear viscosity in the wQGP with 
$\eta/s\simeq 0.8$ for $\alpha_s=0.3$ \cite{Arnold-shear}. Recent calculations 
using Boltzmann approach of multiparton scatterings (BAMPS) show that  
the small shear viscosity over entropy density can also be obtained by considering
perturbative QCD inelastic scattering $gg\rightarrow ggg$,  see Ref. \cite{XuZhe}. 

In fluid dynamics, there is another important transport coefficient, the bulk viscosity
$\zeta$, which has often been neglected in hydrodynamic simulation of nuclear
collisions. The zero bulk viscosity is for a conformal equation of state and also a
reasonable approximation for the weakly interacting gas of quarks and gluons. 
For example, the perturbative QCD calculation gives $\zeta/s=0.02 \alpha_s^2$ for
$0.06<\alpha_s<0.3$ \cite{Arnold-bulk}. However,
recent lattice QCD results show that the bulk viscosity over
entropy density ratio $\zeta/s$ rises dramatically up to the order of $1.0$ near the
critical temperature $T_c$ \cite{LAT-xis-KT,LAT-xis-Meyer}. 
(There are still some subtle issues to determine the bulk viscosity of QCD through 
calculating the correlations of the energy-momentum tensor on the lattice, see more
detailed discussion in Ref. \cite{correlation-Karsch}.)
The sharp peak of bulk
viscosity at $T_c$ has also been observed in the linear sigma model \cite{bulk-Paech-Pratt}
and in the real scalar model \cite{Li-Huang}. The increasing tendency of $\zeta/s$ 
has been shown in a massless pion gas \cite{bulk-Chen} and in the NJL model below $T_c$ 
\cite{Bulk-Sasaki}.
The large bulk viscosity near phase transition is related to the
non-conformal equation of state \cite{LAT-EOS-G, LAT-EOS-Nf2}, and the correlation
between the bulk viscosity and the conformal anomaly has been investigated
in Ref. \cite{Bulk-Nicola}.

The sharp rise of the bulk viscosity will lead to the breakdown of the
hydrodynamic approximation around the critical temperature. 
The effect of large bulk viscosity on hadronization and
freeze-out processes of QGP created at heavy ion collisions has been 
discussed in Refs. \cite{bulk-Mishustin,bulk-Muller,bulk-review-Kharzeev,bulk-kapusta}.
The authors of Ref. \cite{bulk-Mishustin} pointed out the possibility that a sharp
rise of bulk viscosity near phase transition induces an instability in the hydrodynamic
flow of the plasma, and this mode will blow up and tear the system into droplets.
Another scenario is pointed out in Ref. \cite{LAT-xis-KT,bulk-review-Kharzeev} that the large
bulk viscosity near phase transition might induce ``soft statistical hadronization", i.e.
the expansion of QCD matter close to the phase transition is accompanied by the
production of many soft partons, which may be manifested through both a decrease
of the average transverse momentum of the resulting particles and an increase in
the total particle multiplicity.

Due to the complexity of QCD in the regime of strong coupling, results on hot
quark matter from lattice calculation and hydrodynamic simulation
are still lack of analytic understanding. In recent years, the anti-de Sitter/conformal 
field theory (AdS/CFT) correspondence has generated enormous interest in using 
thermal ${\cal N} = 4$ super-Yang-Mills theory (SYM) to understand sQGP.
The shear viscosity to entropy density ratio $\eta/s$ is as small as $1/4\pi$ in the 
strongly coupled SYM plasma \cite{bound}. However, a conspicuous shortcoming 
of this approach is the conformality of SYM: the square of the speed of sound $c_s^2$ 
always equals to $1/3$ and the bulk viscosity is always zero at all temperatures in this 
theory. Though $\zeta/s$ at $T_c$ is non-zero for a class of black hole solutions 
resembling the equation of state of QCD,  the magnitude is less than $0.1$ 
\cite{Gubser-EOS}, which is too small comparing with lattice QCD results.

An alternative nonperturbative approach to study QCD phase transition is by using
effective models. There is still no satisfied dynamic model which can describe
deconfinement phase transition successfully. We thus focus on effective models 
which can describe chiral symmetry restoration. 
QCD with two-flavor massless quarks has a global 
symmetry $SU(2)_R \times SU(2)_L$, which is isomorphic to $O(4)$. The chiral 
symmetry restoration in the case of 2-flavor QCD is of second order phase 
transition, and the universal critical behavior  falls in the 
same universality class as the $O(4)$ model \cite{Pisarski-Wilczek}.
It has been argued that the chiral phase transition at finite chemical 
potential near critical point belongs to the $Z(2)$ universality class 
\cite{Uni-CEP}. This motivates us to study the thermodynamic and transport 
properties of $Z(2)$ and $O(4)$ models near phase transition in this paper. 

The critical phenomena in the $Z(2)$ and $O(4)$ models have been well-known,
and the singular behavior of the static and dynamic properties has been
very well studied. However, due to the finite size and time effects for the
system created in heavy ion collisions, the critical singularity will not
show up in the observables. Therefore, in this paper, we will not focus
on the singularities at the critical point. We will study the thermodynamic
and transport properties in the Hartree approximation of the 
Cornwall-Jackiw-Tomboulis (CJT) formalism \cite{CJT}. At finite temperature, 
the naive perturbative expansion in powers of the coupling constant breaks down, 
and CJT formalism provides a convenient resummation method. 
In the Hartree approximation, we cannot perform the precise critical 
behavior at phase transition, but we can get the qualitative behavior near 
phase transition.

It has been found in Ref. \cite{etas-scalar} that in the simplest 
real scalar model with $Z(2)$ symmetry breaking in the vacuum, $\eta /s$ 
behaves the same way as that in systems of water, helium and nitrogen in 
first-, second-order phase transitions and crossover  \cite{Csernai:2006zz}. 
In Ref. \cite{Li-Huang}, we have investigated the equation of state and bulk 
viscosity in the real scalar model in the case of 2nd order phase transition, 
and we have found that the thermodynamic properties and
transport properties in this simple model at strong coupling are similar to 
those of the complex QCD system. 
In this paper, we will systematically investigate the thermodynamic properties 
and bulk viscosity of the $Z(2)$ model and the $O(4)$ model in the cases 
of first-, second-order phase transitions and crossover in the framework of 
CJT formalism. 

This paper is organized as follows. In Sec. \ref{sec-CJT}, we introduce 
the CJT formalism in the $Z(2)$ and $O(4)$ model. In Sec. \ref{sec-thermo-bulk}, we 
introduce the thermodynamic quantities and the bulk viscosity. 
In Sec. \ref{sec-Numerical}, we present our numerical 
results. At the end, we give discussions and summary in Sec. \ref{sec-Summary}.

\section{$Z(2)$ and $O(4)$ models in the framework of CJT formalism}
\label{sec-CJT}

\subsection{CJT formalism}

At finite temperature $T$, the temperature introduces a new energy scale 
which can conspire with the typical momentum scale $p$ of a process so 
that $gT/p$ is no longer of order $g$, but can be of order $1$ \cite{Braaten}.
Consequently, all terms of order $gT/p$ have to be taken into account
which requires the resummation of certain classes of diagrams. 
The CJT formalism, which is equivalent to the $\Phi$-functional 
approach of Luttinger and Ward \cite{Luttinger} and Baym \cite{Baym},
provides such a convenient resummation method. 
In this paper, we follow the notations used in Ref. \cite{CJT-Dirk}.

The CJT formalism can be viewed as a prescription for computing the 
effective action of a given theory, it generalizes the concept of the 
effective action $\Gamma[\bar{\phi}]$ for the expectation value
$\bar{\phi}$ of the one-point function in the presence of
external sources to that for the effective action $\Gamma[\bar{\phi},\bar{G}]$
for $\bar{\phi}$ and the expectation value $\bar{G}$ of the
two-point function in the presence of external sources, with
\begin{equation} \label{effact}
\Gamma[\bar{\phi},\bar{G}] = \Gamma_0[\bar{\phi}] + \frac{1}{2} \,
{\rm Tr} \ln \bar{G}^{-1}
+ \frac{1}{2} {\rm Tr} ( G_0^{-1} \bar{G} - 1) + \Gamma_2[\bar{\phi}, \bar{G}].
\end{equation}
Here, 
$\Gamma_0[\bar{\phi}]$ is the tree-level action, $G_0^{-1}$ the inverse tree-level
two-point function, and $\Gamma_2[\bar{\phi},\bar{G}]$
the sum of all two-particle irreducible
(2PI) vacuum diagrams with internal lines given by $\bar{G}$. 

The stationary points of this functional,
\begin{equation} \label{stat}
\left.\frac{\delta \Gamma[\bar{\phi},\bar{G}]}{\delta \bar{\phi}}
\right|_{\bar{\phi}=\varphi,\bar{G} = G} = 0\;, \;\;\;\;
\left.\frac{\delta \Gamma[\bar{\phi},\bar{G}]}{\delta \bar{G}}
\right|_{\bar{\phi}=\varphi,\bar{G} = G} = 0\;,
\end{equation}
provide self-consistent
equations for the expectation values of the one- and two-point functions 
$\bar{\phi}$ and $\bar{G}$ in the absence of external sources,
denoted as $\varphi$ and $G$, respectively. The stationarity conditions
Eq. (\ref{stat}) for the effective action are Dyson-Schwinger equations for the
one- and two-point Green's functions of the theory.
The Dyson-Schwinger equation for the two-point Green's function can be derived 
and has the form of
\begin{equation}
 G^{-1} = G_0^{-1} + \Pi\;,
\end{equation}
where
\begin{equation} \label{Pi}
\Pi \equiv - 2 \left. \frac{\delta \Gamma_2 [\bar{\phi},\bar{G}]}{\delta
\bar{G}} \right|_{\bar{\phi}= \varphi, \bar{G} = G}
\end{equation}
is the self-energy. 

In general, the CJT formalism resums one-particle irreducible diagrams
to all orders. As long as $\Gamma_2$ contains all
2PI diagrams, the CJT effective action is exact. However, it is
practically impossible to compute all 2PI diagrams, and one has
to truncate $\Gamma_2$ at some order in the number of loops.
The advantage of the CJT formalism is that any truncation of
$\Gamma_2$ yields a many-body approximation scheme which 
preserves the symmetries of the tree-level action. The solution of 
Eqs.\ (\ref{stat}) is thermodynamically consistent and conserves the 
Noether currents.

The CJT formalism is quite useful for studying
theories with spontaneously broken symmetries.  
In the following, we use the CJT formalism 
studying the $Z(2)$ and $O(4)$ models with
spontaneous symmetry breaking in the vacuum and
symmetry restoration at finite temperature. 

\subsection{$Z(2)$ model in the CJT formalism}

The critical end point has been studied by using the Ginzburg-Landau 
effective potential of the order parameter field (scalar field) up to
the sixth order, e.g. see Ref. \cite{Uni-CEP}. Here we introduce the 
real scalar theory including the sextet interaction which described by the 
Lagrangian
\begin{equation}
\mathcal{L}=\frac{1}{2}(\partial _{\mu }\phi )^{2}-\frac{1}{2}a\phi ^{2}-%
\frac{1}{4}b\phi ^{4}-\frac{1}{6}c\phi ^{6} + H \phi.
\end{equation}%
When $H=0$, this theory is invariant under $\phi \rightarrow -\phi $ and 
has a $Z_{2}$ symmetry. 

Unlike in the Ginzburg-Landau effective potential \cite{Uni-CEP} where
$a,b,c$ are functions of temperature, here $a,b,c$ are model parameters,
which determine the vacuum properties. The system at finite temperature
will be evaluated in the CJT formalism. We will
discuss the following four cases: 1) $c=0, b>0, a>0, H=0$, the system is always in
the symmetric phase. 2) $c=0, b>0, a<0, H=0$, the vacuum at $T=0$ breaks the $%
Z_{2}$ symmetry spontaneously, and the symmetry is restored at higher $%
T $ with a second-order phase transition. 3) $c=0, b>0, a<0, H\neq0$, the $Z(2)$
symmetry is explicitly broken, and the system will experience a crossover
at high temperature. 4) $c>0,b<0, a>0, H=0$, the broken symmetry is restored 
at high $T$ with a first-order phase transition.

Assuming translation invariance, we consider effective potential $\Omega$
instead of effective action $\Gamma$, these two quantities are related via:
\begin{equation}
\Gamma=-\frac{V}{T}\Omega,
\end{equation}
where $V$ is the 3-volume of the system. 
The effective potential in the CJT formalism
reads \cite{CJT-Dirk} 
\begin{eqnarray}
\Omega[\bar{\phi},\bar{G}] =\Omega_0(\bar{\phi})\,+\frac{1}{2}\int_{K}\left[ \,\ln
\bar{G}^{-1}(K)+\bar{G}_{0}^{-1}(K)\,\bar{G}(K)-1\,\right] 
+\,\,\Omega_{2}[\bar{\phi},\bar{G}] ,
\end{eqnarray}%
where 
\begin{equation}
\Omega_0(\bar{\phi})=\frac{a}{2}~\bar{\phi}^{2}+\frac{b}{4}~\bar{\phi}^{4}+\frac{c}{6}~\bar{\phi}%
^{6} -H\bar{\phi}
\end{equation}
is the tree-level potential, and ${\bar G}({\bar G}_{0})$ is the full(tree-level)
propagator: 
\begin{equation}
\bar{G}^{-1}(K,\bar{\phi})=-K^{2}+m^{2}(\bar{\phi})\;,\newline
\bar{G}_{0}^{-1}(K,\bar{\phi})=-K^{2}+m_{0}^{2}(\bar{\phi})\;,
\end{equation}%
with the tree-level mass $m_{0}^{2}=a+3b~\bar{\phi}^{2}+5c~\bar{\phi}^{4}$.
In Hartree approximation, the 2PI potential $\Omega_{2}$ only includes 
\begin{equation}
\Omega_{2}[\bar{\phi},{\bar G}]=\left( \frac{3}{4}b+\frac{15}{2}c\bar{\phi}^{2}\right)
\left(\int_K {\bar G}(K)\right) ^{2}+\frac{15}{6}c \left( \int_K {\bar G}(K) \right)^{3},
\label{V2}
\end{equation}%
The self-consistent one- and two-point Green's functions satisfy 
\begin{equation}
\left. \frac{\delta \Omega}{\delta \bar{\phi}}\right\vert _{\bar{\phi}=\phi
,{\bar G}=G}\equiv 0\;,\;\;\;\left. \frac{\delta \Omega}{\delta {\bar G}}%
\right\vert _{\bar{\phi}=\phi,{\bar G}=G}\equiv 0\;\;\;\;.
\end{equation}%
This allows us to solve $\phi$ and $m$ through the coupled Dyson-Schwinger equations: 
\begin{gather}
a\phi+b\phi^{3}+c\phi^{5}+(3b\phi+10c\phi^{3})\int_K G(K) +15c\phi\left(\int_K G(K)\right)^{2} =H\,,  \notag \\
m^{2}-m_{0}^{2}=3(b+10c\phi^{2}) \int_K G(K)+15c \left( \int_K G(K) \right)^{2}.\,
\end{gather}

\subsection{$O(4)$ model in the CJT formalism} 

The Lagrangian of the $O(N)$ model reads 
\be \label{L} {\cal L}(\phi) =
\frac{1}{2} \,
\partial_\mu{\mbox{\boldmath$\phi$}}\cdot\partial^\mu
\mbox{\boldmath$\phi$} - \frac{a}{2}
\mbox{\boldmath$\phi$}\cdot\mbox{\boldmath$\phi$} - \frac{b}{4N}
(\mbox{\boldmath$\phi$}\cdot\mbox{\boldmath$\phi$})^2
+H\phi_1\,\,\,\, , \ee 
where $\mbox{\boldmath $\phi$}\equiv(\phi_1,...,\phi_N)$ is an $O(N)$ vector.  
We identify the first component $\phi_1$ with the
$\sigma$ field and the remaining $N-1$ components as the $\pi$
fields. The last term $H \phi_1$ breaks the symmetry 
explicitly and has been introduced in order
to generate masses for the pions.   For $H=0$,
$a>0$ and $b>0$, the Lagrangian is invariant under $O(N)$ rotations
of the fields. For $H=0$, $a<0$ and $b>0$, this symmetry
is spontaneously broken down to $O(N-1)$, leading to $N-1$ Goldstone
bosons (the pions), and the field \mbox{\boldmath $\phi$} obtains a non-vanishing
vacuum expectation value $\bar{\phi}$. 
The $O(N)$ symmetry will be restored at finite
temperature with a second-order phase transition.
By shifting the field as $ \mbox{\boldmath
$\phi$}\rightarrow\mbox{\boldmath $\phi$}+\bar{\phi}$,  the
``classical potential" takes the form 
\bea
\label{classicalpotential} 
\Omega_0(\bar{\phi})
 & = & \frac{a}{2} \bar{\phi}\,^{2} +
        \frac{b}{4N} \bar{\phi}\,^{4} -H \bar{\phi}\,\,\,\,. 
\eea
The inverse tree-level propagator which corresponds to the above
Lagrangian density is 
\bea 
\label{Dsigma}
{\bar G}^{-1}_{0\sigma}(k;\bar{\phi}) &=& -k\,^{2} + a +
        \frac{3\, b}{N}\, \bar{\phi}\,^{2} \,\,\,\,\,\, , \\
{\bar G}^{-1}_{0{\pi}}(k;\bar{\phi}) &=& -k\,^{2} + a +
        \frac{b}{N}\, \bar{\phi}\,^{2} \,\,\,\,\,\, ,
\label{Dpi}\eea
where $G^{-1}_{0\sigma}$,\,$G^{-1}_{0\pi}$ are sigma and pion
propagator respectively.

The CJT effective potential of the $O(N)$ model can be written 
as the following function of full propagators:

\bea
\label{CJT-potential}
\Omega(\bar{\phi},\bar{G}_{\sigma},\bar{G}_{\pi})
 & = & \Omega_0(\bar{\phi}) + 
   \half \int_k \, \left[ \ln {\bar G}^{-1}_{\sigma}(k) + 
        {\bar G}_{0\sigma}^{-1}(k;\bar{\phi})\, \bar{G}_{\sigma}(k)-1 \right] \nonumber \\
 & & + \frac{N-1}{2} \int_k \, \left[ \ln {\bar G}^{-1}_{\pi}(k) \,+
         {\bar G}^{-1}_{0{\pi}}(k;\bar{\phi})\, \bar{G}_{\pi}(k)-1\right]\, 
 +\Omega_{2}(\bar{\phi},\bar{G}_{\sigma},\bar{G}_{\pi})\,\,\,\, . 
\eea
In the Hartree approximation, the 2PI effective potential $\Omega_2$ takes the form of
\bea
\Omega_2=(N+1)(N-1)\, \frac{b}{4 N} \left[
        \int_Q \, \bar{G}_{\pi}(Q)\right]^{2}+ 3\,\frac{b}{4N}
        \left[ \int_Q \, \bar{G}_{\sigma}(Q) \right]^{2}+ 2\,(N-1) \frac{b}{4N}
        \int_Q\, \bar{G}_{\sigma}(Q) 
        \int_L\, \bar{G}_{\pi}(L) \,\,.
\eea

The stationarity condition is written as
\bea 
\label{stationvphi} 
\left. \frac{\delta
\Omega[\bar{\phi},\bar{G}_{\sigma},\bar{G}_{\pi}]}{\delta \bar{\phi}} \,
\right|_{\bar{\phi}=\phi,\,\bar{G}_{\sigma}=G_{\sigma},\,
\bar{G}_{\pi}= G_{\pi}}= 0 \,\,\,\,\,\,\, ,\\
\label{stationvGsig}
\left. \frac{\delta
\Omega[\bar{\phi},\bar{G}_{\sigma},\bar{G}_{\pi}]}{\delta
\bar{G}_{\sigma}(k)} \,
\right|_{\bar{\phi}=\phi,\,\bar{G}_{\sigma}=
G_{\sigma},\,\bar{G}_{\pi}
=G_{\pi}}= 0 \,\,\,\,\,\,\, ,\\
\label{stationvGpi}\left. \frac{\delta
\Omega[\bar{\phi},\bar{G}_{\sigma},\bar{G}_{\pi}]}{\delta
\bar{G}_{\pi}(k)} \,
\right|_{\bar{\phi}=\phi,\,\bar{G}_{\sigma}=
{G}_{\sigma},\,\bar{G}_{\pi}=G_{\pi}}= 0 \,\,\,\,\,\,\,\, ,
\label{stationvG} \eea 
which determines the expectation values of
the one- and two-point functions in the absence of external sources
$\phi$ and $G_{\sigma}$, $G_{\pi}$. Similarly, the
Schwinger--Dyson equations for sigma and pion propagators
 for the effective potential $\Omega$ are given by
\bea\label{sda}
G_{\sigma}^{-1}(k) & = &
G_{0\sigma}^{-1}(k;\phi) +
\Sigma_{\sigma}(k)\, = -k^{2} + M^{2}_{\sigma} \,\,,\\
G_{\pi}^{-1}(k) & = & G_{0\pi}^{-1}(k;\phi) +
\Sigma_{\pi}(k)\, = - k^{2} + M^{2}_{\pi} \,\,. \label{sdb}
\eea 
By solving Eq.(\ref{stationvGsig}) and (\ref{stationvGpi}), the 
corresponding self-energy can be obtained as 
\bea 
\label{selfenergy2}
\Sigma_{\sigma}(k) \equiv && 2 \left. \frac{\delta \Omega_2
[\bar{\phi},\bar{G}_{\sigma},\bar{G}_{\pi}]}{\delta
\bar{G}_{\sigma}(k)}
\right|_{\bar{\phi}=\phi,\,\bar{G}_{\sigma}=G_{\sigma},\,
\bar{G}_{\pi}=G_{\pi}} \,\,\,\,\,\,\,\,\,\,\,\,\,,\\
\Sigma_{\pi}(k) \equiv && \frac{2}{N-1} \left. \frac{\delta \Omega_2
[\bar{\phi},\bar{G}_{\sigma},\bar{G}_{\pi}]}{\delta
\bar{G}_{\pi}(k)}
\right|_{\bar{\phi}=\phi,\,\bar{G}_{\sigma}=G_{\sigma},\,
\bar{G}_{\pi}=G_{\pi}} \,. \eea

In our following numerical calculations, we will take the case of $N=4$. 

\section{The equation of state, sound velocity, specific heat, interaction measure and bulk viscosity}

\label{sec-thermo-bulk}

Once we have the effective potential $\Omega$, we can derive all thermodynamic properties 
of the system. The entropy density is determined by taking the derivative of effective 
potential with respect to the temperature, i.e, 
\be
\label{entropy}
s=-\partial \Omega(\phi)/\partial T .
\ee
In the symmetry breaking case, the vacuum effective potential or the vacuum energy density
is negative, i.e, 
\be
\label{vacuum-energy}
\Omega_{v}=\Omega(\phi)|_{T=0}<0.
\ee
As the standard treatment in lattice calculation, we introduce the normalized 
pressure density $p_T$ and energy density $\epsilon_T$ as 
\be
p_T=-\Omega_T, \,\, \epsilon_T=-p_T+ T s,
\ee
with 
\be
\Omega_T=\Omega(\phi)-\Omega_{v}.
\ee
The equation of state $p_T(\epsilon_T)$ is
an important input into hydrodynamics.
The square of the speed of sound $C_s^2$ is related to $p_T/\epsilon_T$ and has the form of
\begin{equation}
C_s^2=\frac{{\rm d}p}{{\rm d}\epsilon}=\frac{s}{T {\rm d}s/{\rm d}T}=\frac{s}{C_v},
\end{equation}
where 
\be
C_v=\partial \epsilon/\partial T,
\ee
is the specific heat.
At the critical temperature, the entropy density as well as
the energy density change most quickly with temperature, thus one expect that $C_s^2$
should have a minimum at $T_c$.
The trace anomaly of the energy-momentum tensor ${\cal T}^{\mu\nu}$
\begin{equation}
\Delta=\frac{{\cal T}^{\mu\mu}}{T^4}\equiv \frac{\epsilon_T-3 p_T}{T^4}
=T\frac{\partial}{\partial T}(p_T/T^4)
\end{equation}
is a dimensionless quantity, which is also called the "interaction measure".


The bulk viscosity is related to the correlation function of the trace of the
energy-momentum tensor $\theta^\mu_\mu$:
\begin{equation}
\label{kubo}
\zeta = \frac{1}{9}\lim_{\omega\to 0}\frac{1}{\omega}\int_0^\infty dt \int d^3r\,e^{i\omega t}\,\langle [\theta^\mu_\mu(x),\theta^\mu_\mu(0)]\rangle \,.
\end{equation}
According to the result derived from low energy theorem, in the low frequency region, 
the bulk viscosity takes the form of ~\cite{LAT-xis-KT}
\begin{eqnarray}\label{ze}
\,\zeta &=& \frac{1}{9\,\omega_0}\left\{ T^5\frac{\partial}{\partial T}\frac{(\epsilon_T-3p_T)}{T^4}
+16|\epsilon_v|\right\}\,, \nonumber \\
 & = & \frac{1}{9\,\omega_0} \left\{- 16 \epsilon_T+9 T S + T C_v + 16 |\epsilon_v| \right\}\,.
\end{eqnarray}
with the negative vacuum energy density $\epsilon_v=\Omega_v=\Omega(\phi)|_{T=0}$,
and the parameter $\omega_0 = \omega_0(T)$ is a scale at which the perturbation theory
becomes valid. From the above formula, we can see that the bulk viscosity is proportional
to the specific heat $C_v$ near phase transition, thus $\zeta/s$ behaves as $1/C_s^2$ near 
$T_c$ in this appproximation.

\section{Thermodynamic properties of $Z(2)$ and $O(4)$ models}
\label{sec-Numerical} 

In this section, we show our numerical results for thermodynamic properties of
the $Z(2)$ and $O(4)$ models. 

\subsection{$Z(2)$ model without symmetry breaking in the vacuum} 

\begin{figure}[thbp]

\epsfxsize=7.5 cm \epsfysize=6.5cm \epsfbox{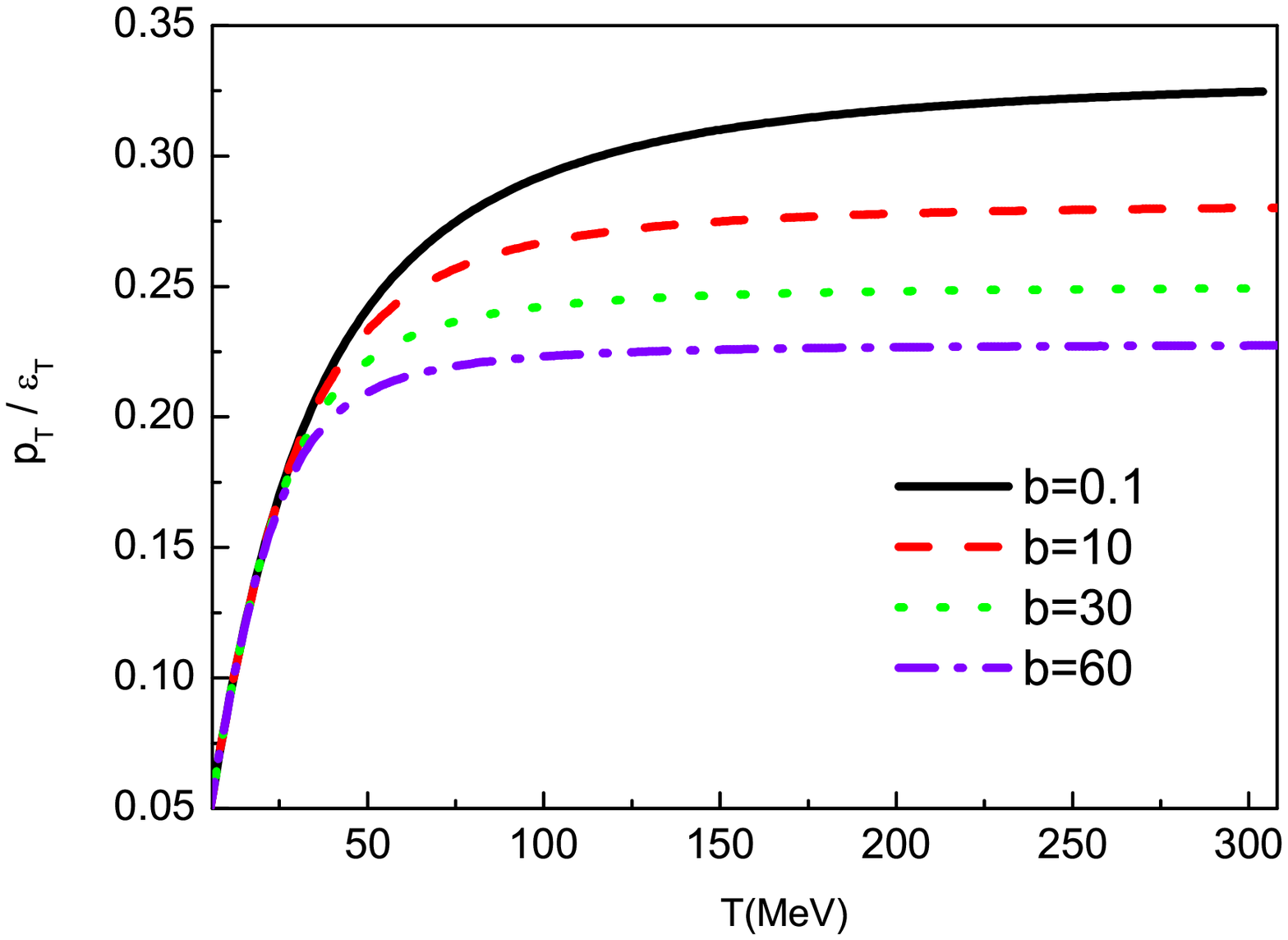}\hspace*{0.1cm}
\epsfxsize=7.5 cm \epsfysize=6.5cm \epsfbox{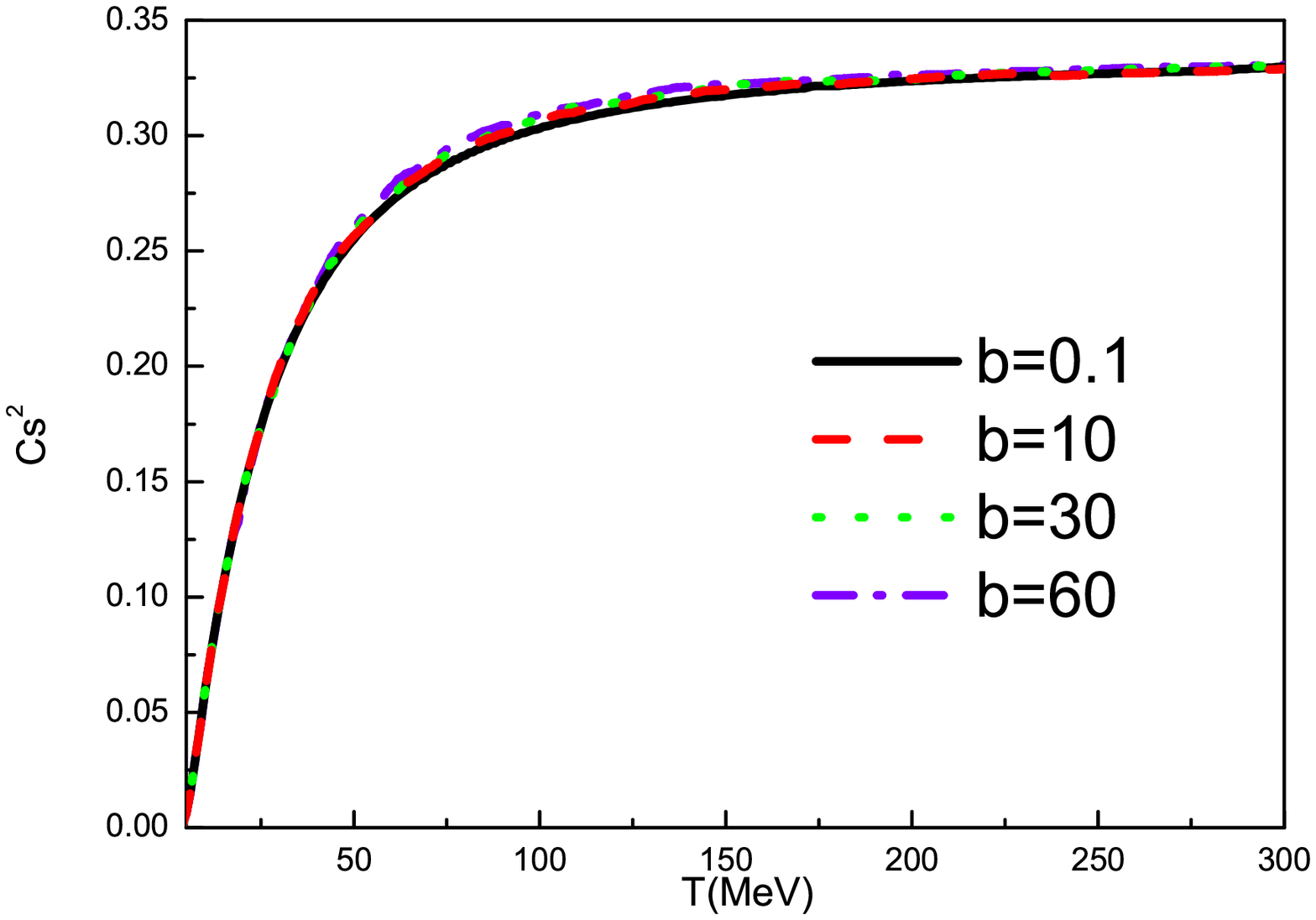}
\vskip -0.05cm \hskip 0.15 cm \textbf{( a ) } \hskip 6.5 cm \textbf{( b )} \\
\epsfxsize=7.5 cm \epsfysize=6.5cm \epsfbox{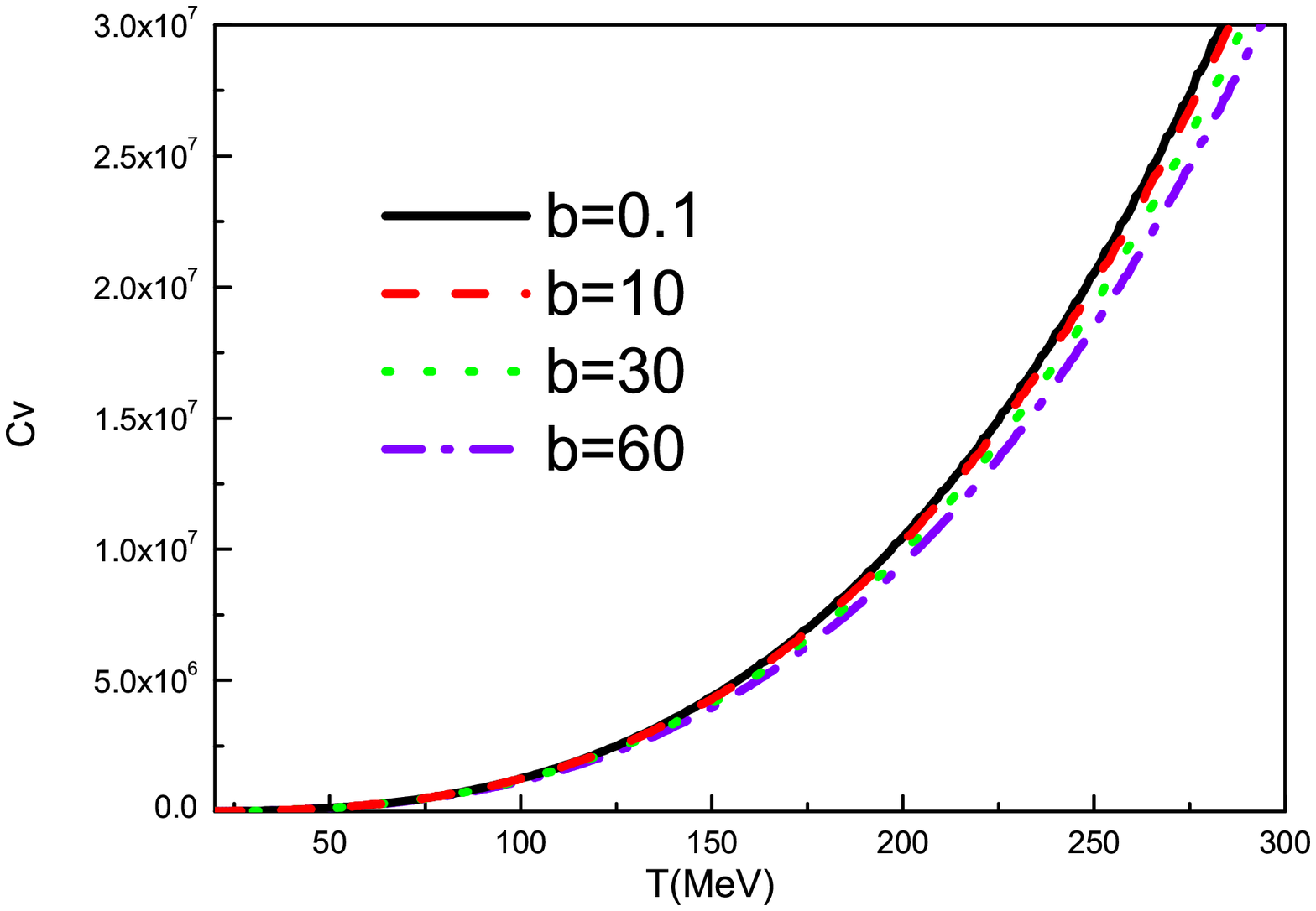} \hspace*{0.1cm}
\epsfxsize=7.5 cm \epsfysize=6.5cm \epsfbox{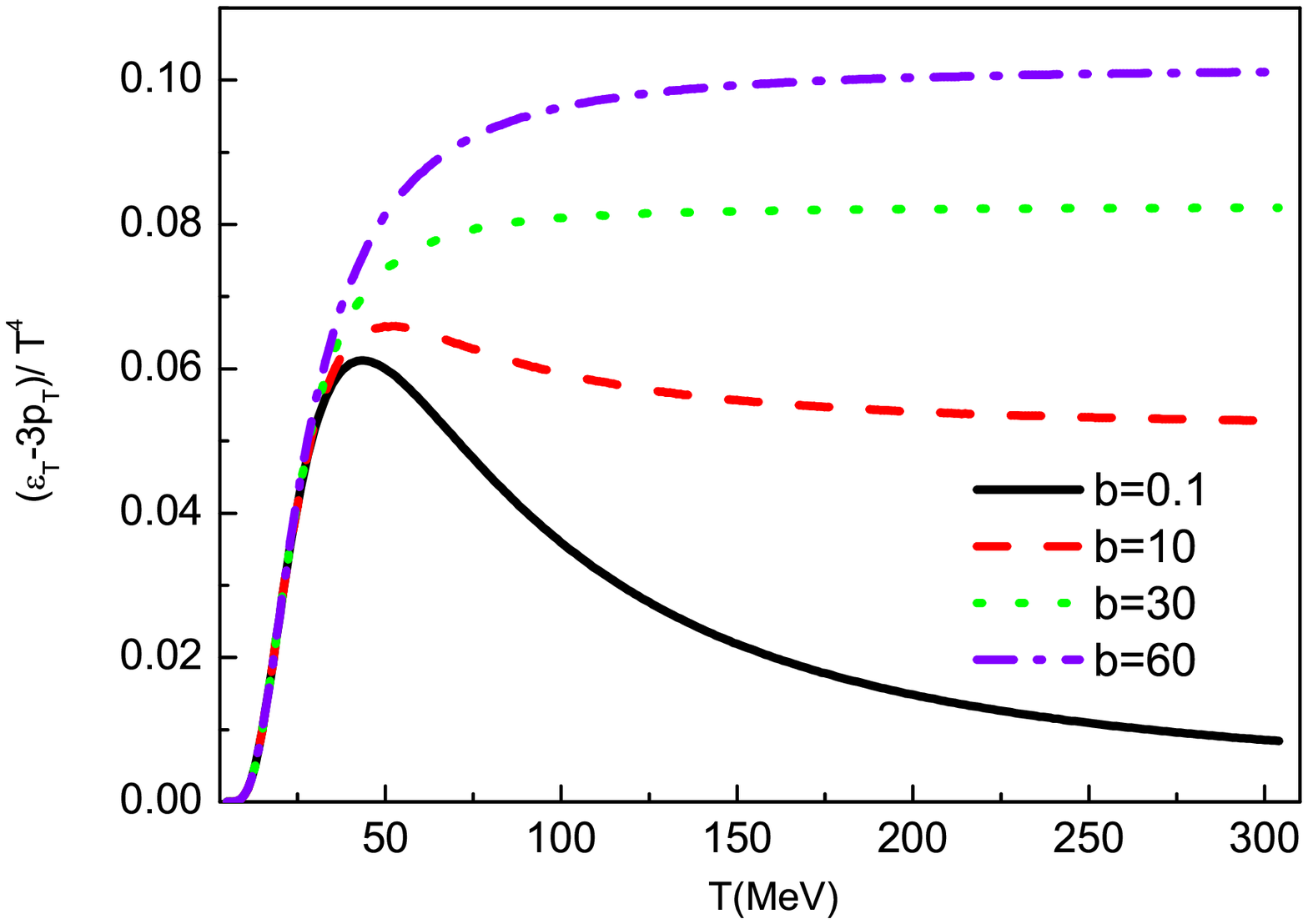}
\vskip -0.05cm \hskip 0.15 cm \textbf{( c ) } \hskip 6.5 cm \textbf{( d )} \\
\epsfxsize=7.5 cm \epsfysize=6.5cm \epsfbox {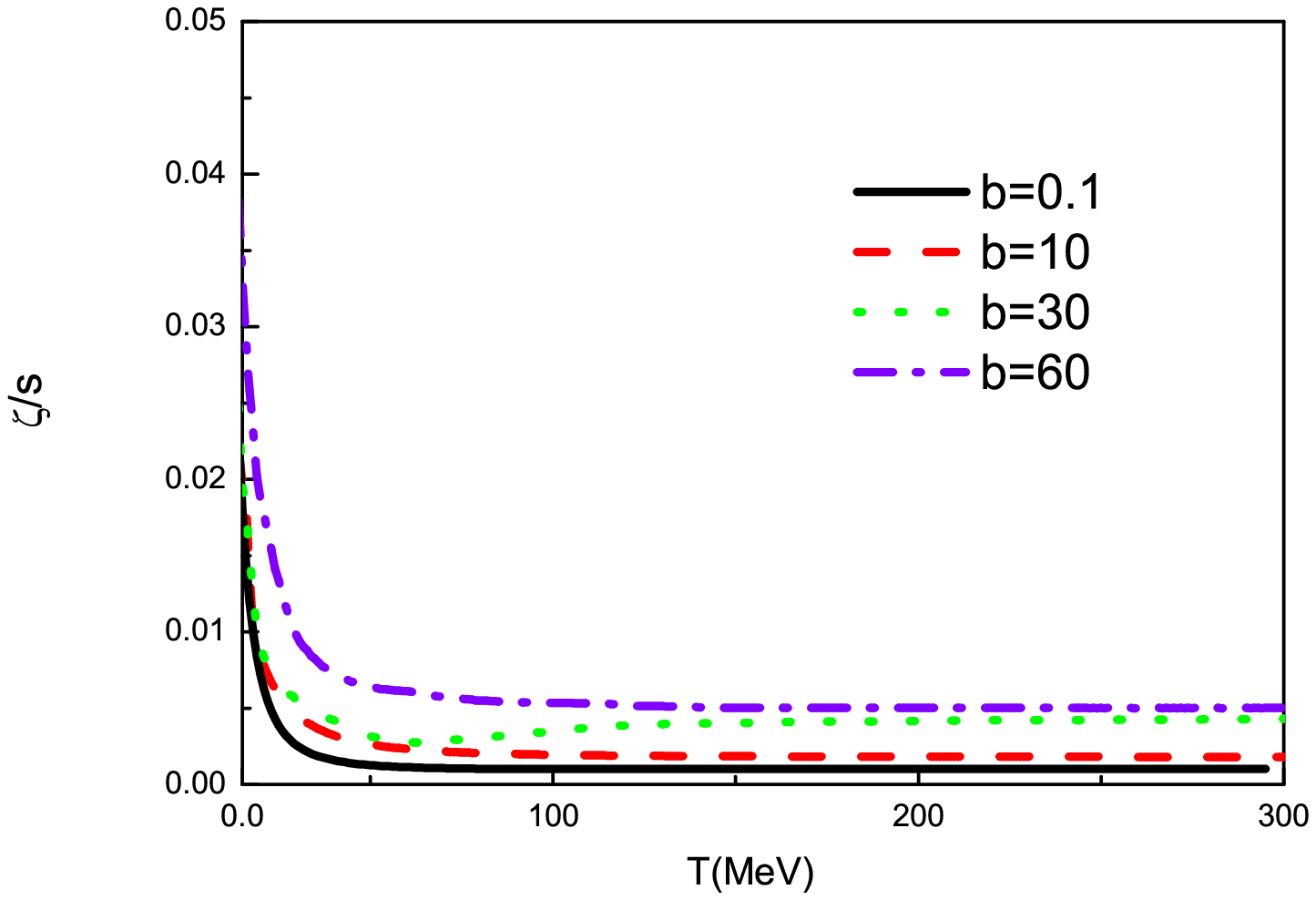} 
\vskip -0.05cm \hskip 0.15 cm \textbf{( e ) } 
 \caption{The ratio of pressure density over energy density $p_T/\epsilon_T$, the 
trace anomaly $(\epsilon_T-3 p_T)/T^4$, the sound velocity square $C_s^2$, 
the specific heat $C_v$ and the bulk viscosity over entropy density ratio $\zeta/s$
as functions of the temperature $T$ for the real scalar model in the symmetric phase.
} 
\label{Symm_N=1}
\end{figure}

We firstly consider the real scalar model without symmetry breaking in the vacuum, i.e. 
$a>0$. In Fig. \ref{Symm_N=1} $(a)-(e)$, we show the ratio of the pressure density over 
energy density $p_T/\epsilon_T$, the trace anomaly $(\epsilon_T-3 p_T)/T^4$,
the square of sound velocity $C_s^2$, the specific heat $C_v$ and the bulk viscosity
over entropy density ratio $\zeta/s$ as functions of the 
temperature $T$ for different coupling strength $b$. The parameters
taken for calculations are: $1)\, a=1000 \,{\rm MeV}^2, \,b=0.1$, 
$2)\, a=10000\, {\rm MeV}^2, \, b=10$, 
$3)\, a=10000\, {\rm MeV}^2, \, b=30$, and $4)\, a=10000 \,{\rm MeV}^2, \, b=60$.
 
In the weak coupling case when $b=0.1$, it is found that the pressure density over energy 
density $p_T/\epsilon_T$, the sound velocity square $C_s^2$ and the specific heat $C_v$ 
show similar behavior as in the ideal gas.  Both $p_T/\epsilon_T$ and $C_s^2$ increase 
with the temperature $T$, and reach the conformal value $1/3$ at high temperature. 
The specific heat  $C_v$ monotonically increases with temperature. There is no much
information about the trace anomaly $(\epsilon_T-3 p_T)/T^4$ for an ideal gas in the
literature. It is found that the trace anomaly $(\epsilon_T-3 p_T)/T^4$ shows a peak at 
low temperature, then decreases monotonically and reaches the conformal value $0$ at high 
temperature. 

The peak of the trace anomaly $(\epsilon_T-3 p_T)/T^4$ at low temperature, which is not 
related to the phase transition was also observed in Ref. \cite{Li-Huang} in the real scalar model
with 2nd-order phase transition and in Ref. \cite{Bulk-Nicola} in the Chiral Perturbation Theory
for the pion gas. In Ref. \cite{Bulk-Nicola}, the low-T peak of the trace anomaly was interpreted
as the explicit conformal breaking, whose contribution comes from massive pions. However, for
the real scalar system, there are no massive pions, it is not clear for us what is the reason 
inducing the conformal symmetry breaking at low temperature.

In the case of strong coupling, we observe that the pressure density over energy 
density $p_T/\epsilon_T$, the sound velocity square $C_s^2$ increase with temperature
and saturate at high temperature. The pressure density over energy density $p_T/\epsilon_T$
saturates at a value smaller than $1/3$, the stronger the coupling strength is, the smaller value
$p_T/\epsilon_T$ saturates. The sound velocity square $C_s^2$ still saturates at $1/3$.  
The specific heat $C_v$ increases with temperature. However, it is found that the low-T peak 
of the trace anomaly slowly disappears with the increase of the coupling strength, and 
at high $T$, the trace anomaly goes to a larger value for stronger coupling strength $b$,
which indicates that the conformal symmetry is broken at high $T$ due to the bare strong 
coupling. Therefore, the trace anomaly at high temperature can indicate the strength
of the coupling, in this sense, the trace anomaly is also called the "interaction measure".
 
In both weak coupling and strong coupling, the bulk viscosity over entropy density $\zeta/s$ decreases
monotonically with the increase of the temperature. At high $T$, $\zeta/s$ reaches its conformal
value $0$ in the case of weak coupling, and reaches a finite value in the case of strong coupling.
However, we don't observe the correlation between the trace anomaly and the bulk viscosity 
as shown in Ref. \cite{Bulk-Nicola}, where it shows that the peak at low temperature in the 
trace anomaly also appears in the bulk viscosity.

\subsection{$Z(2)$ model with a second-order phase transition}

\begin{figure}[thbp]
\epsfxsize=7.5 cm \epsfysize=6.5cm \epsfbox{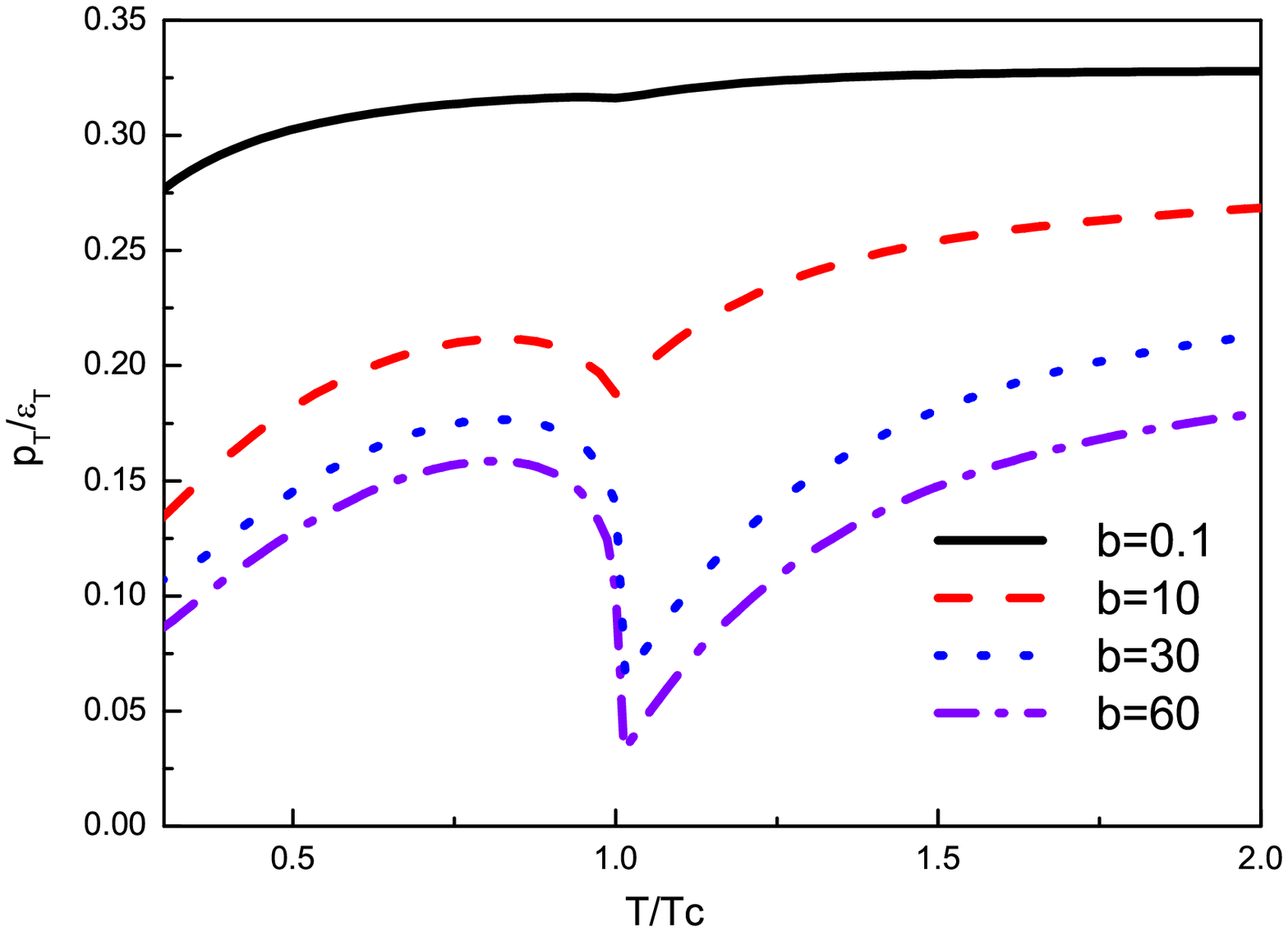}\hspace*{0.1cm}
\epsfxsize=7.5 cm \epsfysize=6.5cm \epsfbox{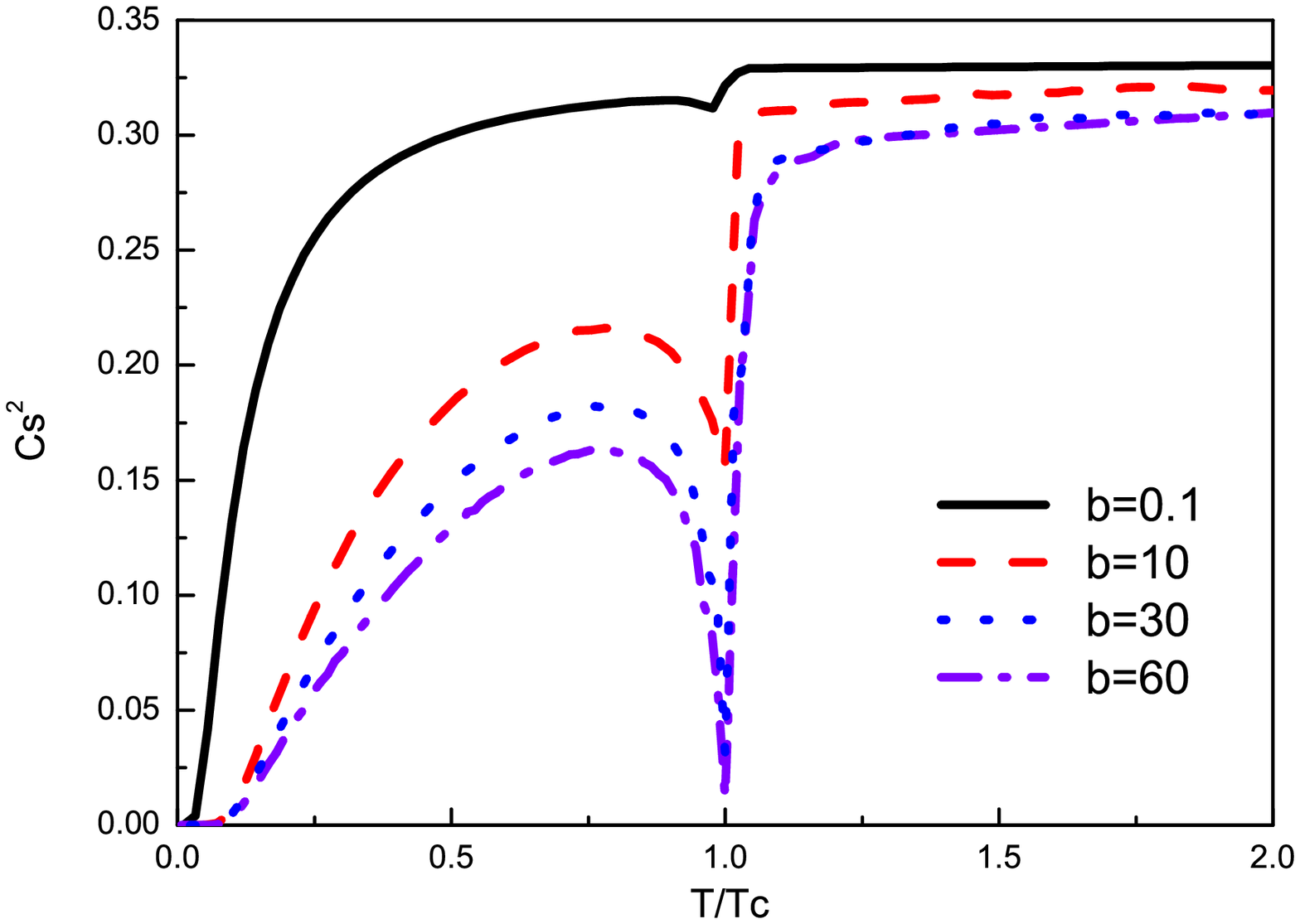}
\vskip -0.05cm \hskip 0.15 cm \textbf{( a ) } \hskip 6.5 cm \textbf{( b )} \\
\epsfxsize=7.5 cm \epsfysize=6.5cm \epsfbox{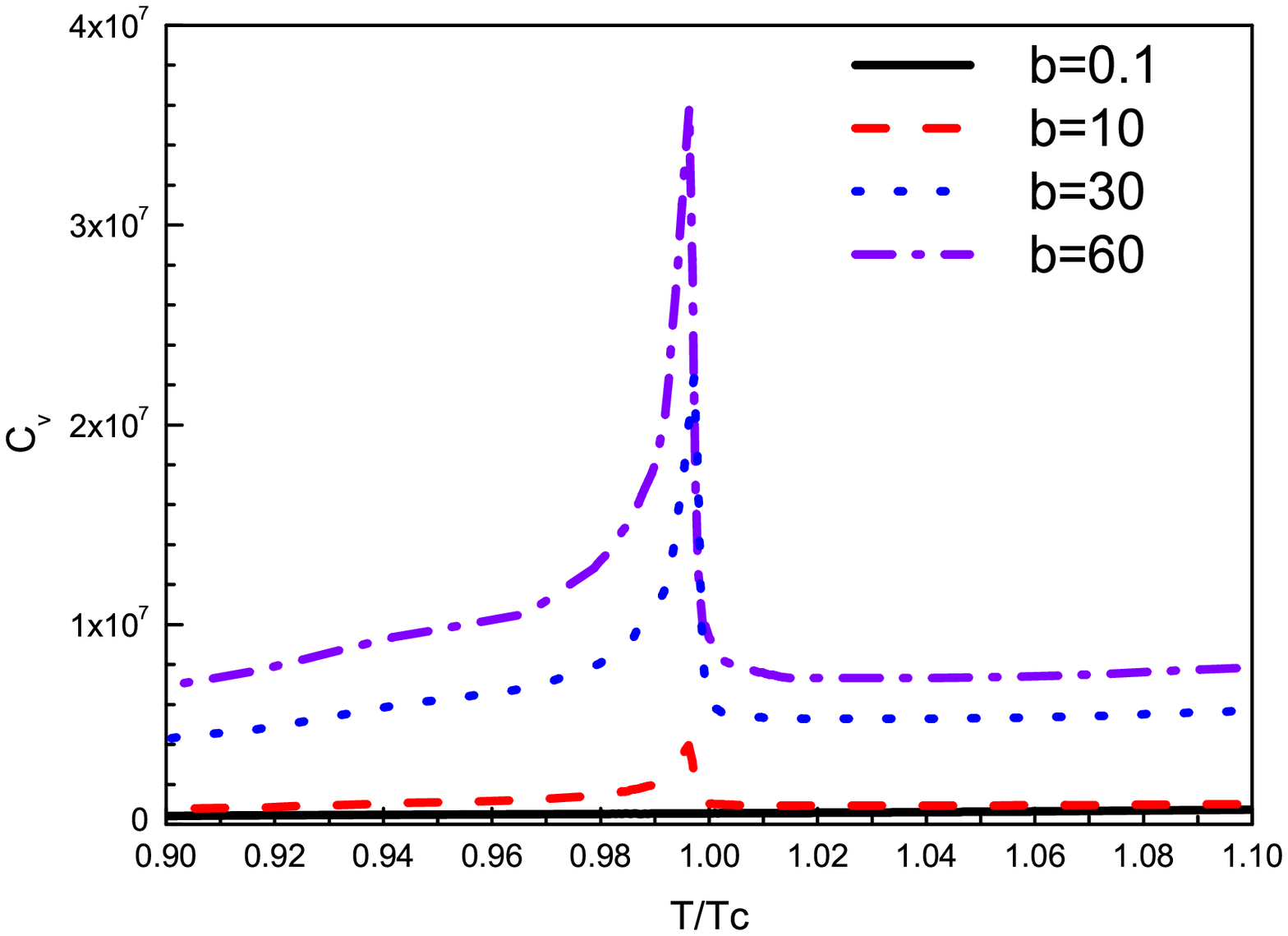} \hspace*{0.1cm}
\epsfxsize=7.5 cm \epsfysize=6.5cm \epsfbox{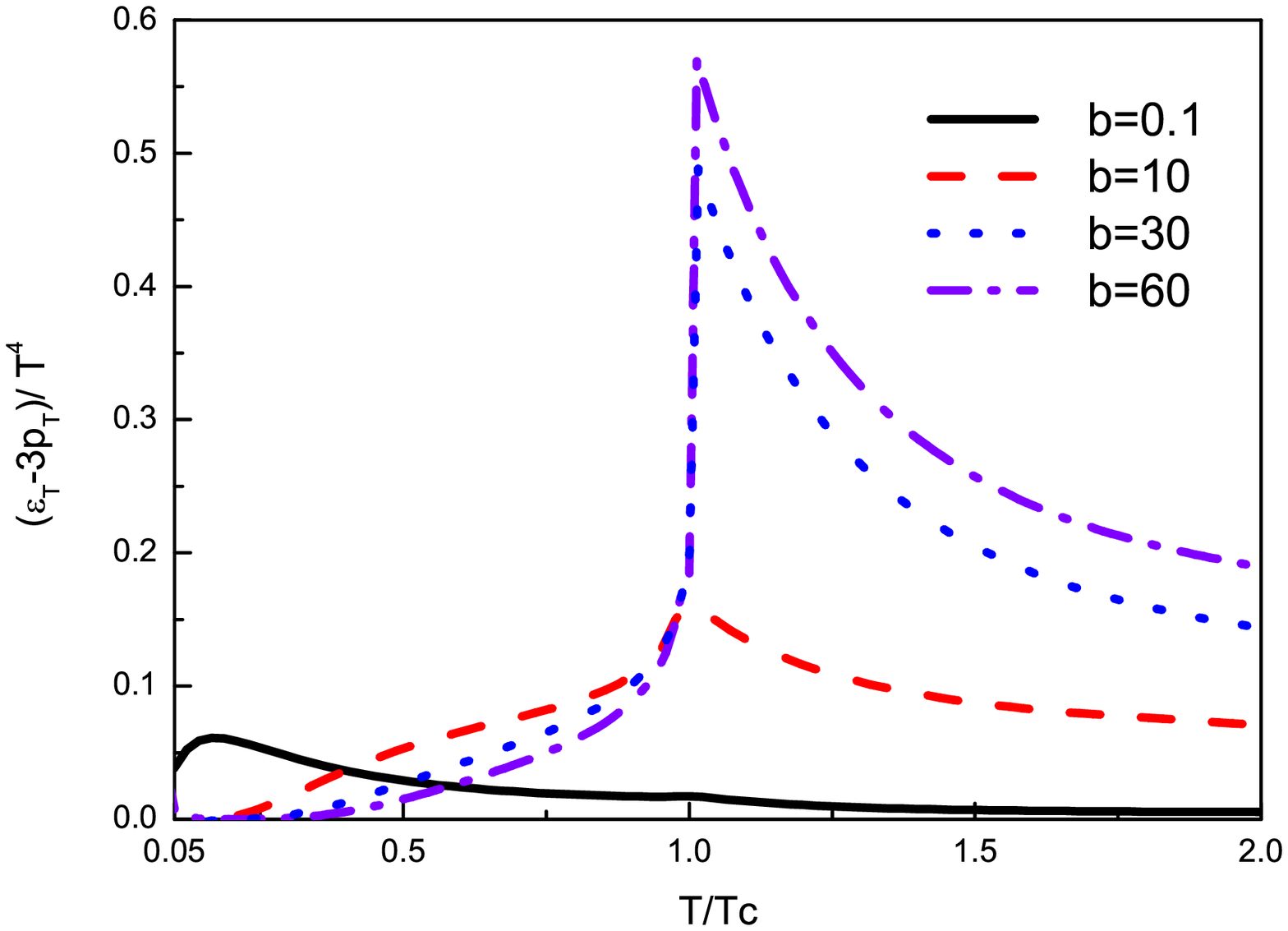}
\vskip -0.05cm \hskip 0.15 cm \textbf{( c ) } \hskip 6.5 cm \textbf{( d )} \\
\epsfxsize=7.5 cm \epsfysize=6.5cm \epsfbox {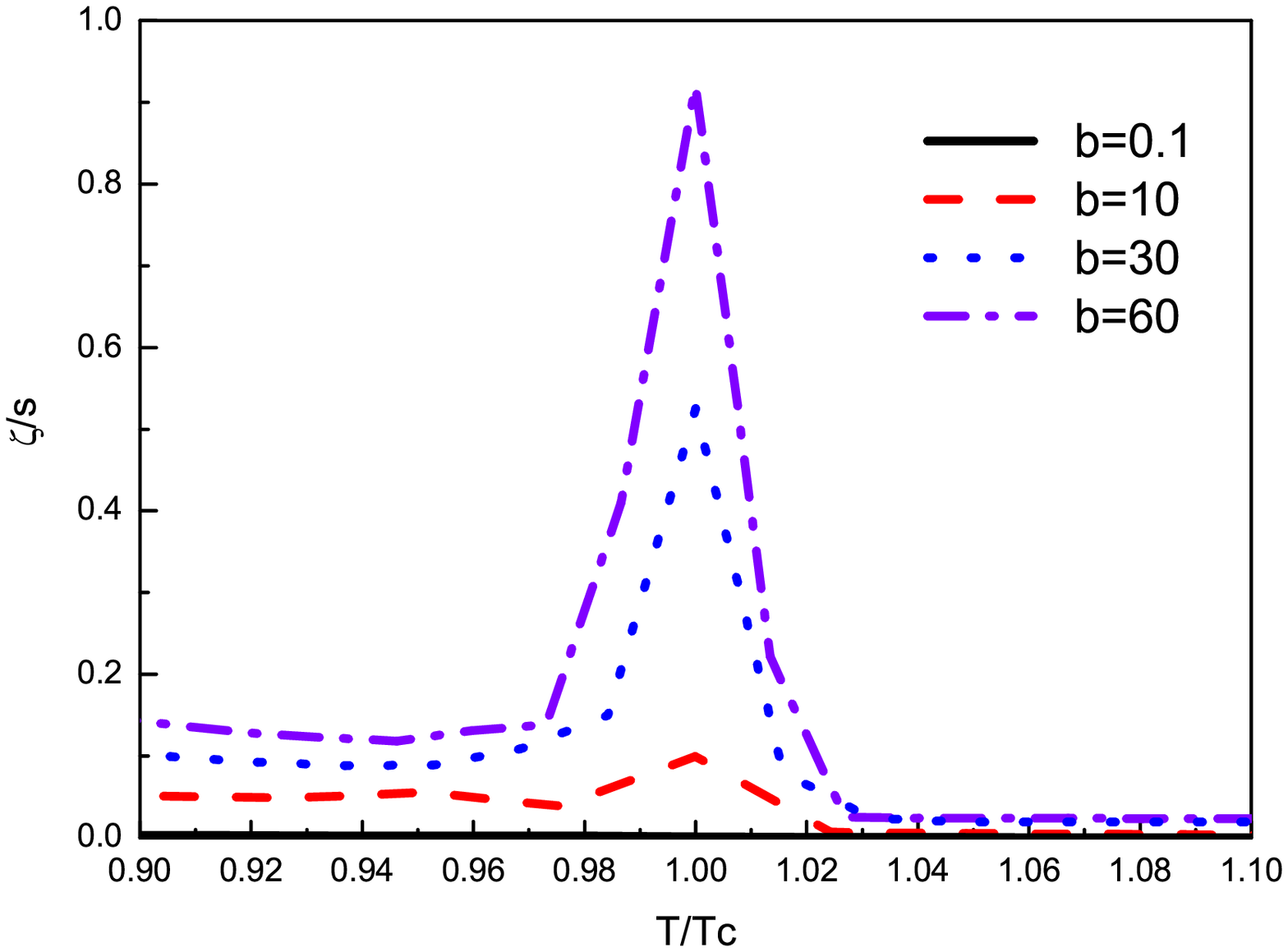} 
\vskip -0.05cm \hskip 0.15 cm \textbf{( e ) } 
 \caption{The ratio of pressure density over energy density $p_T/\epsilon_T$, 
the trace anomaly $(\epsilon_T-3 p_T)/T^4$, the sound velocity square
$C_s^2$, the specific heat $C_v$ and the bulk viscosity over entropy density 
ratio $\zeta/s$ as functions of the temperature $T$ with a second-order 
phase transition in the real scalar model.} \label{2nd_N=1}
\end{figure}

In Fig. \ref{2nd_N=1} $(a)-(e)$, we show the ratio of pressure density over energy density 
$p_T/\epsilon_T$, the trace anomaly $(\epsilon_T-3 p_T)/T^4$,
the sound velocity square $C_s^2$, the specific heat $C_v$ and the bulk viscosity
over entropy density $\zeta/s$ as functions of the temperature $T$ for different coupling 
strength $b$ for the real scalar model with a second order phase transition. The parameters
used for calculations and the corresponding critical temperatures are: 
$1)\, a=-100 \,{\rm MeV}^2,\, b=0.1,\, T_c=88 \,{\rm MeV}$, 
$2)\, a=-10000 \,{\rm MeV}^2,\, b=10,\, T_c=82 \,{\rm MeV}$,
$3)\, a=-50000 \,{\rm MeV}^2,\, b=30,\, T_c=124 \,{\rm MeV}$, 
$4)\, a=-100000 \,{\rm MeV}^2,\, b=60,\, T_c=144 \,{\rm MeV}$.
The main results in this case have been shown in Ref. \cite{Li-Huang}.

In the weak coupling case when $b=0.1$, it is found that the pressure density over energy 
density $p_T/\epsilon_T$, the sound velocity square $C_s^2$ and the specific heat $C_v$ 
show similar behavior as in the ideal gas except near the phase transition region.  
Both $p_T/\epsilon_T$ and $C_s^2$ show a downward cusp at $T_c$, and reach the 
conformal value $1/3$ at high temperature. The specific heat $C_v$ 
exhibits a small upward cusp at $T_c$. In the weak coupling case, we observe double 
peak in the trace anomaly, one smooth peak shows up at low temperature, another upward
cusp appears at $T_c$. 

In the case of strong coupling, at high temperature region $T>T_c$, the behavior of the
the pressure density over energy density $p_T/\epsilon_T$, the sound velocity square 
$C_s^2$, the specific heat $C_v$, and the trace anomaly show similar behavior
as those in the symmetric case. $p_T/\epsilon_T$ and $C_s^2$ increase with temperature 
and saturate at high temperature. The pressure density over energy density $p_T/\epsilon_T$
saturates at a value smaller than $1/3$, the stronger the coupling strength is, the smaller 
saturation value $p_T/\epsilon_T$ will be. The sound velocity square $C_s^2$ still saturates 
at $1/3$. The specific heat $C_v$ increases with temperature. The trace anomaly decreases 
with temperature and goes to a larger value for stronger coupling strength $b$. 

Near 
phase transition region $T\simeq T_c$, both the pressure density over energy density 
$p_T/\epsilon_T$ and the sound velocity square $C_s^2$ show a downward cusp at $T_c$, 
the specific heat $C_v$ and the trace anomaly show an upward cusp at $T_c$. 
When the coupling strength increases, the depth of the downward cusp for the pressure 
density over energy density $p_T/\epsilon_T$ and the sound velocity square $C_s^2$ 
become deeper and deeper, while the height of the upward cusp for the specific heat 
$C_v$ and the trace anomaly becomes higher and higher.

At low temperature region $T<T_c$, both the pressure density over energy density 
$p_T/\epsilon_T$ and the sound velocity square $C_s^2$ show a bump, i.e. $p_T/\epsilon_T$ 
and $C_s^2$ firstly increase with $T$ then decrease with $T$.
However, it is found that the low-T peak of the trace anomaly slowly disappears with 
the increase of the coupling strength.
  
The bulk viscosity over entropy density $\zeta/s$ decreases with $T$
at low temperature region, then rises up at the critical temperature $T_c$,
and decreases further in the temperature $T>T_c$.

\subsection{$Z(2)$ model with explicit symmetry breaking}

\begin{figure}[thbp]

\epsfxsize=7.5 cm \epsfysize=6.5cm \epsfbox{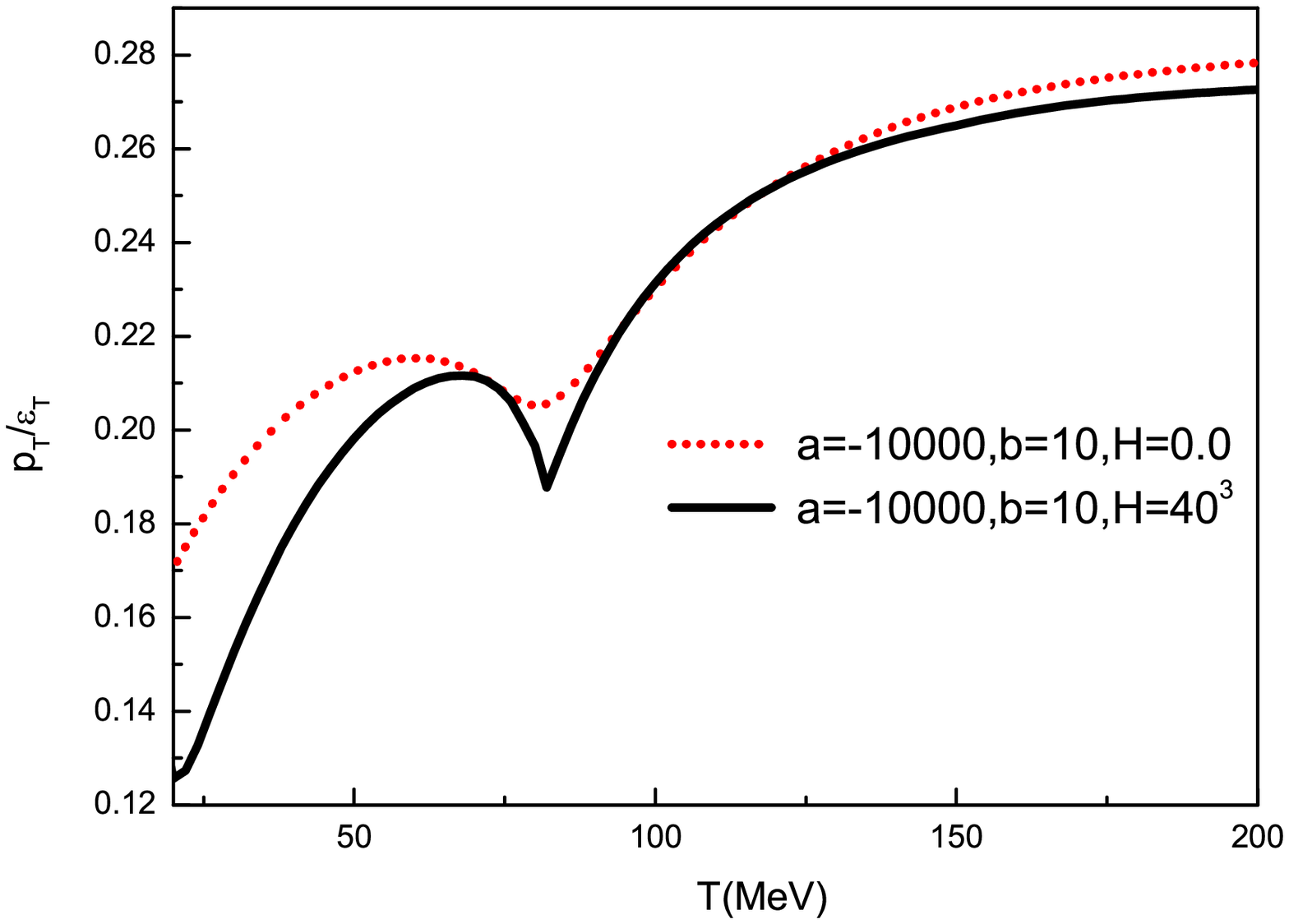}\hspace*{0.1cm}
\epsfxsize=7.5 cm \epsfysize=6.5cm \epsfbox{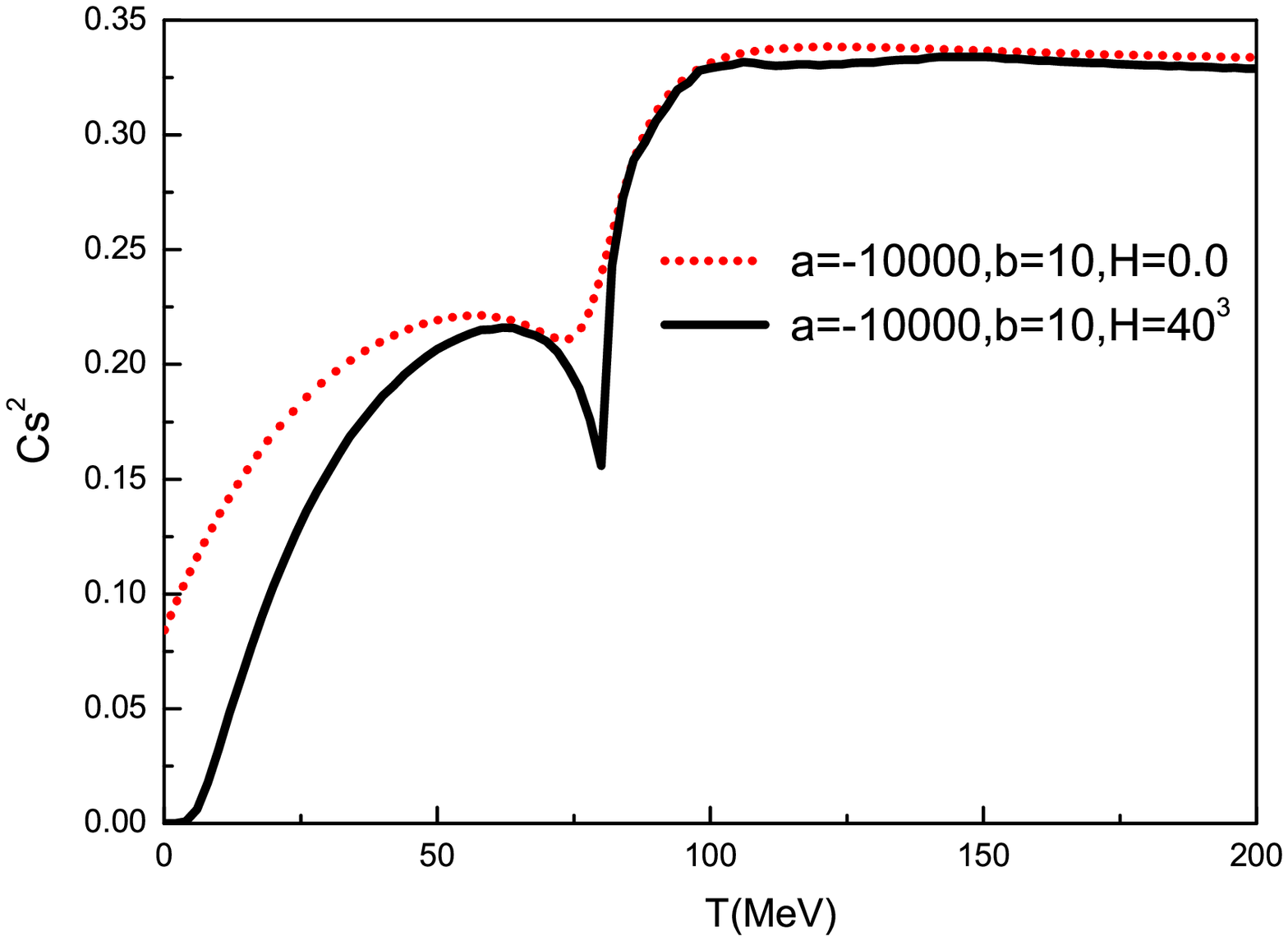}
\vskip -0.05cm \hskip 0.15 cm \textbf{( a ) } \hskip 6.5 cm \textbf{( b )} \\
\epsfxsize=7.5 cm \epsfysize=6.5cm \epsfbox{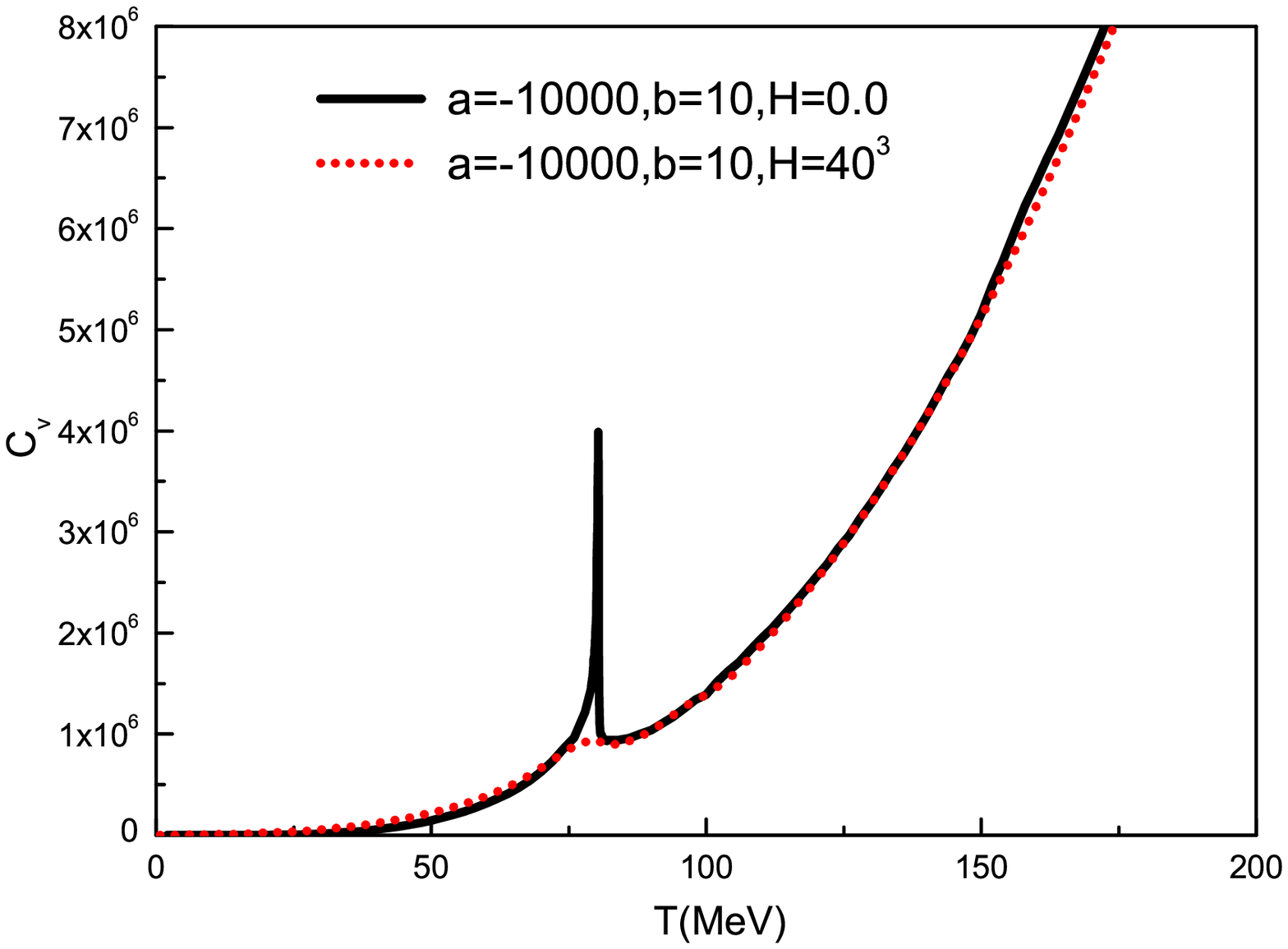}\hspace*{0.1cm}
\epsfxsize=7.5 cm \epsfysize=6.5cm \epsfbox{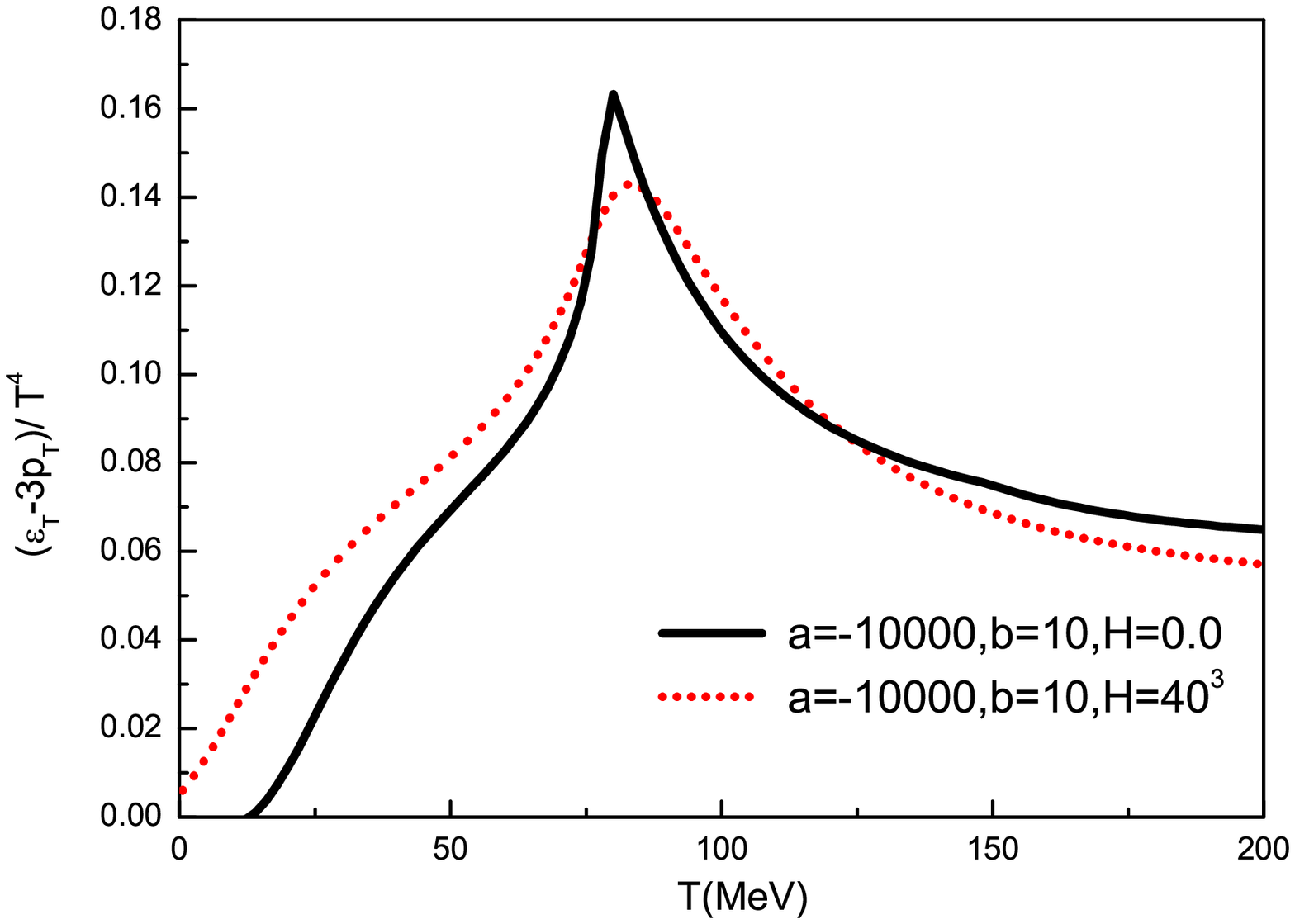}
\vskip -0.05cm \hskip 0.15 cm \textbf{( c ) } \hskip 6.5 cm \textbf{( d )} \\
\epsfxsize=7.5 cm \epsfysize=6.5cm \epsfbox {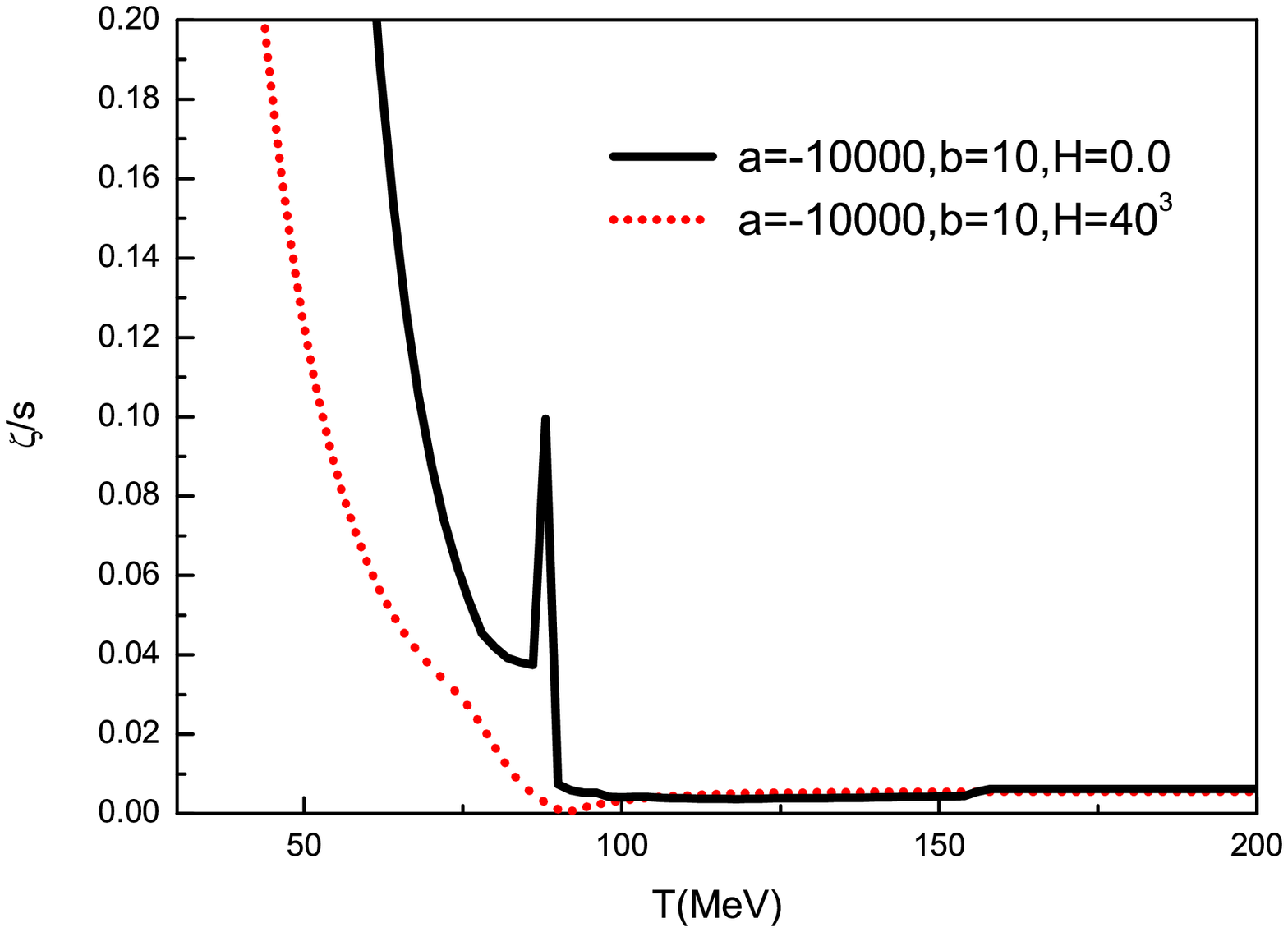} 
\vskip -0.05cm \hskip 0.15 cm \textbf{( e ) } 
\caption{The ratio of pressure density over energy density $p_T/\epsilon_T$, the 
trace anomaly $(\epsilon_T-3 p_T)/T^4$, the sound velocity square $C_s^2$, 
the specific heat $C_v$ and the bulk viscosity over entropy density ratio $\zeta/s$ 
as functions of the temperature $T$ with a crossover and with a second-order phase 
transition in the real scalar model.} 
\label{Cross_N=1}
\end{figure}

In Fig. \ref{Cross_N=1} $(a)-(e)$, we show the ratio of pressure density over 
energy density $p_T/\epsilon_T$, the sound velocity square $C_s^2$,
the trace anomaly $(\epsilon_T-3 p_T)/T^4$, the specific heat $C_v$ and the bulk viscosity
over entropy density $\zeta/s$ as functions of the temperature $T$ for second order
phase transition and crossover for the real scalar model. The parameters used 
for calculations are: $1)\, a=-10000\, {\rm MeV}^2, \, b=10, \, H=0$, and
$2)\, a=-10000 \, {\rm MeV}^2,\, b=10,\, H=(40 {\rm MeV})^3$. The only difference
of the two set of parameters is the value of $H$. When $H=0$, the system experiences
a second order phase transition, and when $H \neq 0$, the system experiences
a crossover.

At both low temperature region $T<T_c$ and high temperature region $T>T_c$, the 
behavior of all the thermodynamic quantities and bulk viscosity don't show much 
difference for these two set of parameters. However, near critical temperature
region $T\simeq T_c$, it is observed all the cusp behaviors are washed out, e.g. 
the downward cusp in the pressure density over 
energy density $p_T/\epsilon_T$ and the sound velocity square $C_s^2$
develops into a shallow valley, and the upward cusp in the trace anomaly 
develops into a smooth peak. Especially, the upward cusp
for the specific heat $C_v$ and the bulk viscosity over entropy density
$\zeta/s$ vanishes and there is no obvious change near critical temperature.  

    
\subsection{$Z(2)$ model with a first-order phase transition}

\begin{figure}[thbp]

\epsfxsize=7.5 cm \epsfysize=6.5cm \epsfbox{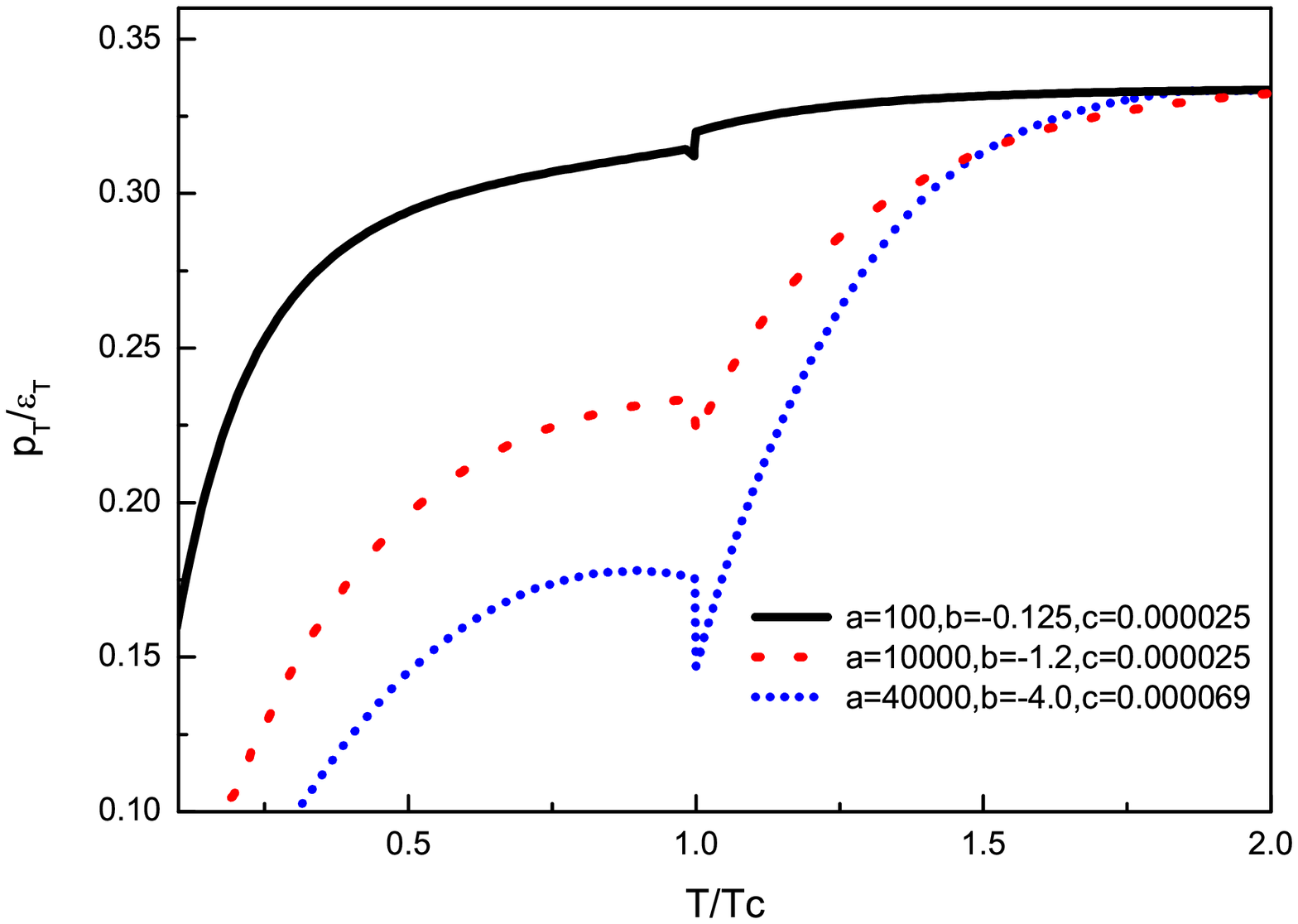}\hspace*{0.1cm}
\epsfxsize=7.5 cm \epsfysize=6.5cm \epsfbox{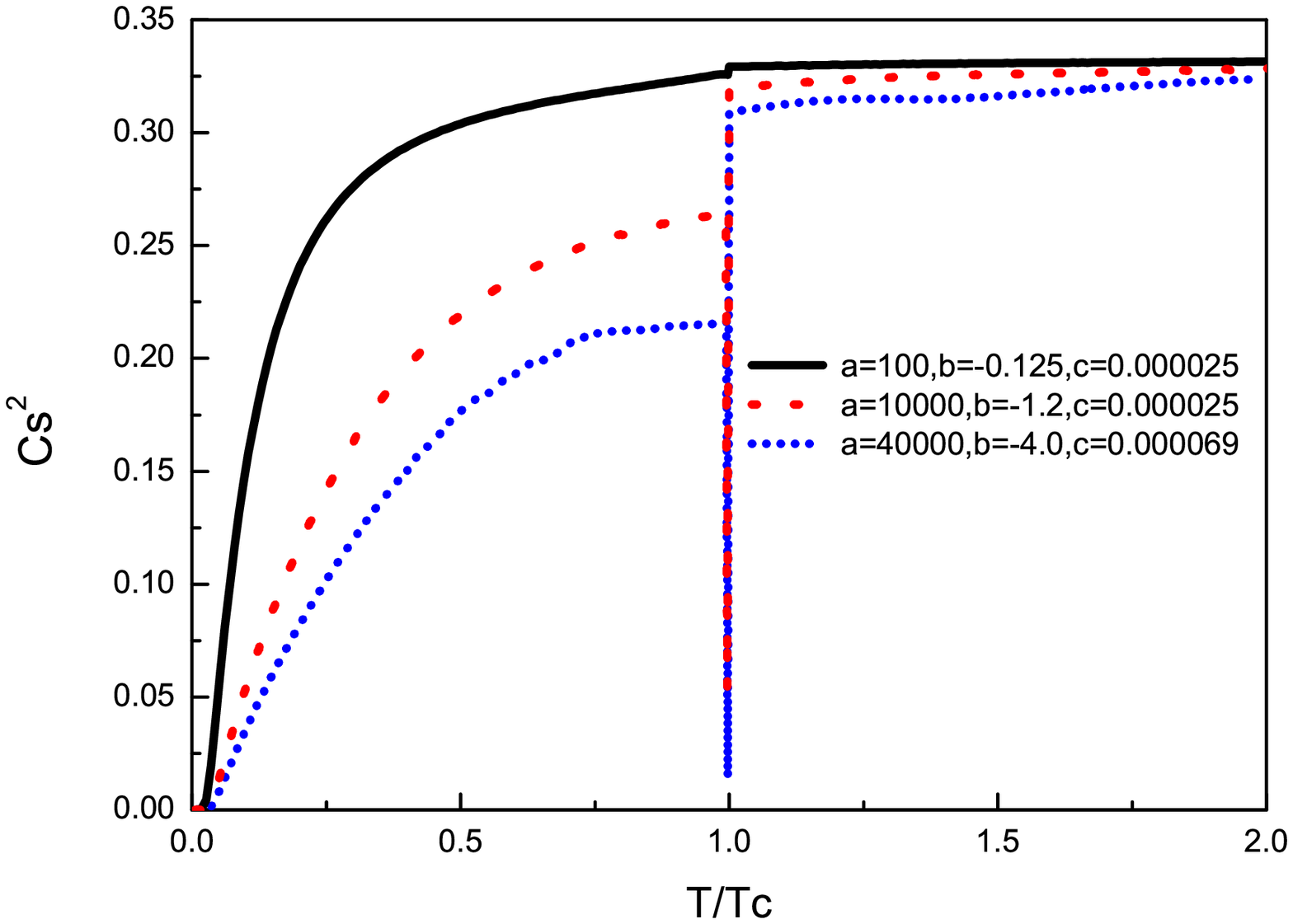}
\vskip -0.05cm \hskip 0.15 cm \textbf{( a ) } \hskip 6.5 cm \textbf{( b )} \\
\epsfxsize=7.5 cm \epsfysize=6.5cm \epsfbox{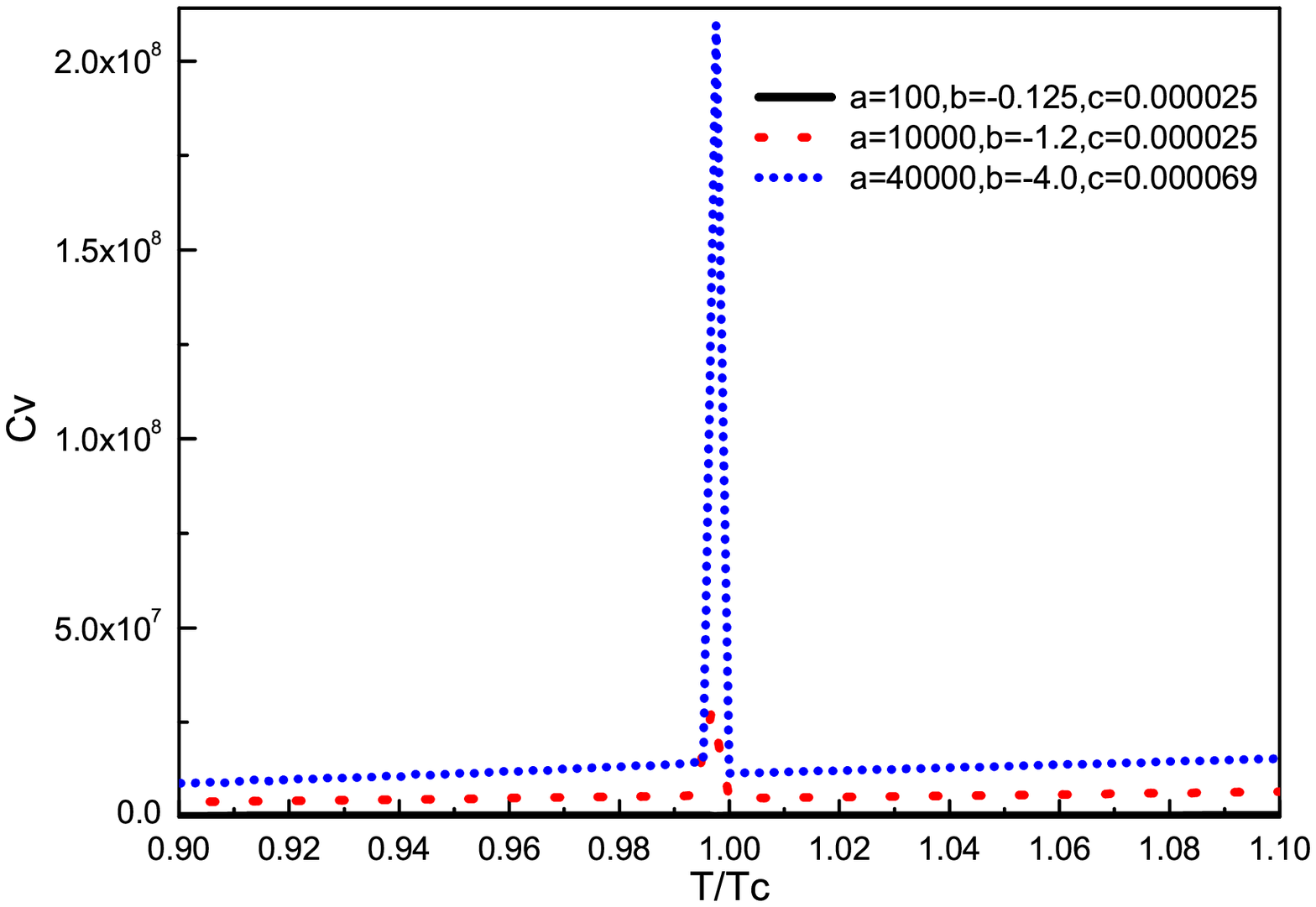} \hspace*{0.1cm}
\epsfxsize=7.5 cm \epsfysize=6.5cm \epsfbox{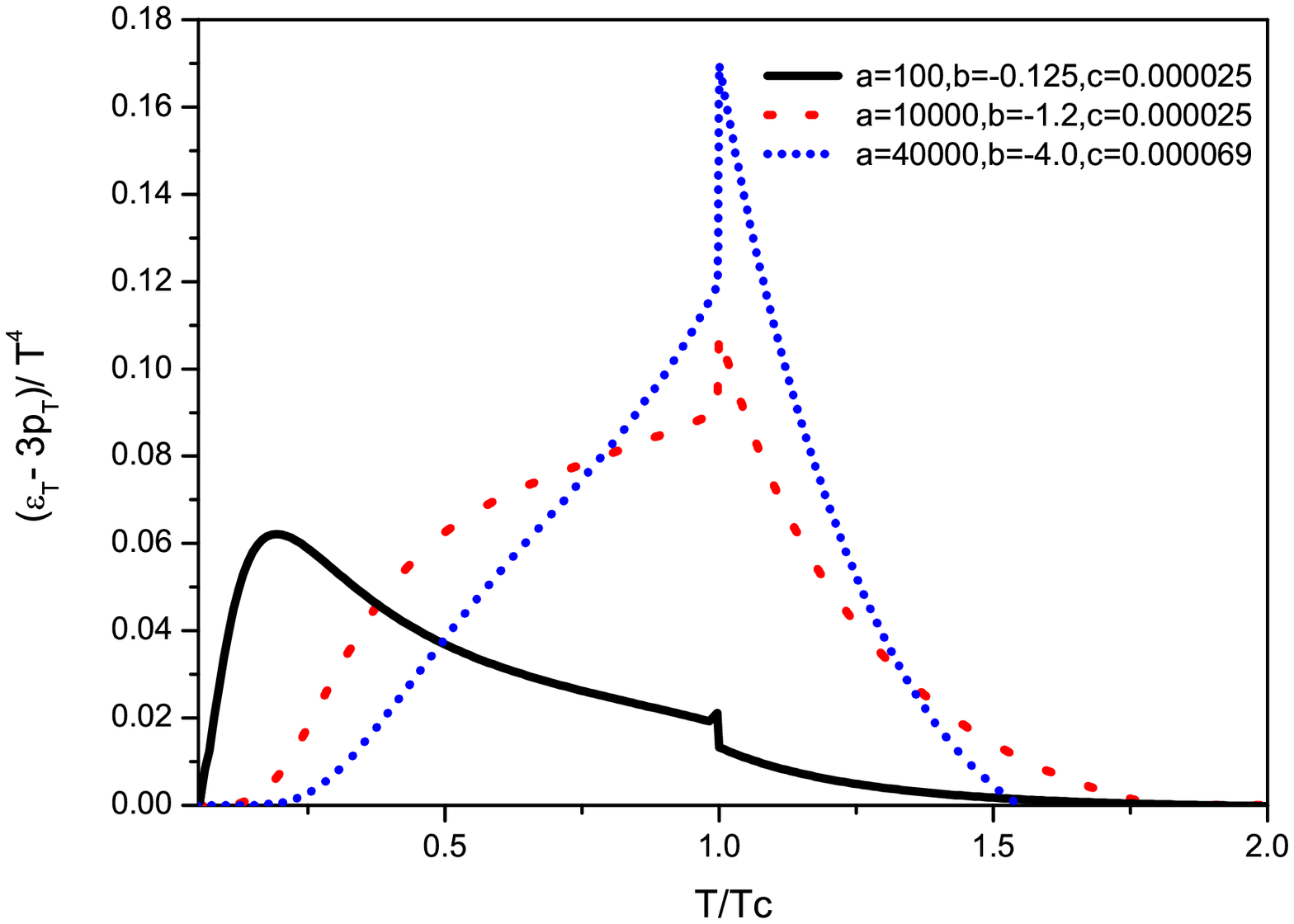}
\vskip -0.05cm \hskip 0.15 cm \textbf{( c ) } \hskip 6.5 cm \textbf{( d )} \\
\epsfxsize=7.5 cm \epsfysize=6.5cm \epsfbox {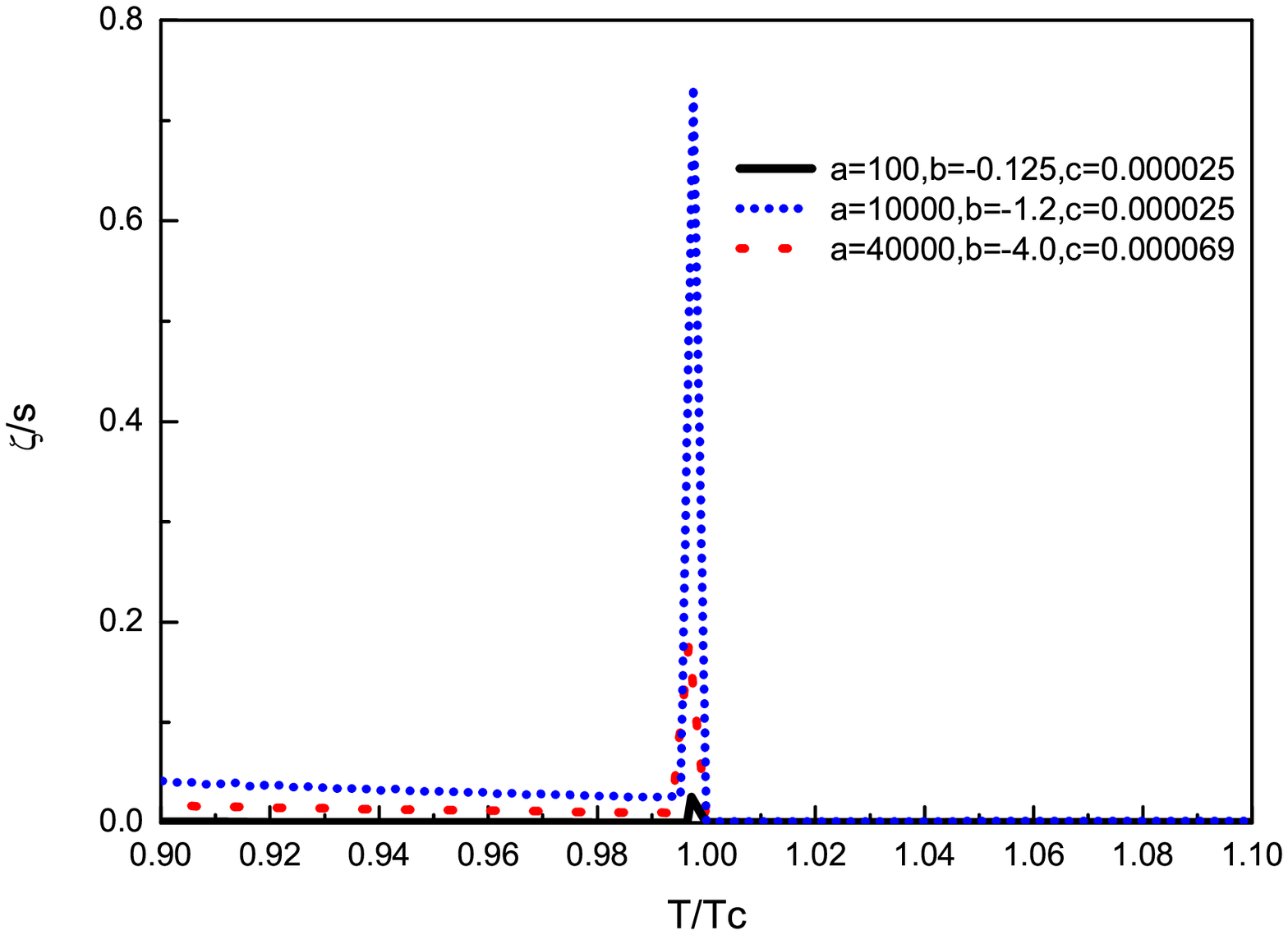} 
\vskip -0.05cm \hskip 0.15 cm \textbf{( e ) } 

\caption{The ratio of pressure density over energy density $p_T/\epsilon_T$, the 
trace anomaly $(\epsilon_T-3 p_T)/T^4$, the sound velocity square $C_s^2$, 
the specific heat $C_v$ and the bulk viscosity over entropy density ratio $\zeta/s$
as functions of the temperature $T$ with a first-order phase transition in the
real scalar model.} 
\label{1st_N=1}
\end{figure}

In Fig. \ref{1st_N=1} $(a)-(e)$, we show the ratio of pressure density over energy density 
$p_T/\epsilon_T$, the sound velocity square $C_s^2$, the trace anomaly 
$(\epsilon_T-3 p_T)/T^4$, the specific heat $C_v$ and the bulk viscosity
over entropy density $\zeta/s$ as functions of the temperature $T$ 
for the real scalar model with a first order phase transition. 
The parameters used for calculations and corresponding critical temperatures
are: $1)\, a=100\, {\rm MeV}^2,\, b=-0.125,\,
c=0.000025 \,{\rm MeV}^{-2}, \, T_c=59 ~{\rm MeV}$,
$2)\,a=10000 \,{\rm MeV}^2,\, b=-1.2, \, c=0.000025 \, {\rm MeV}^{-2},\, 
T_c=155\, {\rm MeV}$, $3)\, a=40000\, {\rm MeV}^2,\, b=-4,\, c=0.000069 \,{\rm MeV}^{-2},
\, T_c=204 \,{\rm MeV}$.

At high temperature region $T>T_c$, the behavior of the
pressure density over energy density $p_T/\epsilon_T$, the sound velocity square 
$C_s^2$, the specific heat $C_v$, and the trace anomaly show similar behavior
as those in the case of symmetric phase. $p_T/\epsilon_T$ and $C_s^2$ 
increase with temperature and saturate at the conformal value $1/3$ 
at high temperature. The specific heat $C_v$ increases with temperature 
and the trace anomaly decreases with temperature and goes to the conformal value $0$.

At low temperature region $T<T_c$, the behavior of the pressure density over energy density 
$p_T/\epsilon_T$, the sound velocity square $C_s^2$, the specific heat $C_v$, and the trace 
anomaly also show similar behavior as those in the case of symmetric phase. 
Both the pressure density over energy density $p_T/\epsilon_T$ and the sound velocity square 
$C_s^2$ monotonically increase with the temperature. The low-T peak of the trace anomaly 
still shows up in the weak coupling case and slowly disappears with 
the increase of the coupling strength.

Near phase transition region $T\simeq T_c$, the behavior of the pressure density over energy density 
$p_T/\epsilon_T$, the sound velocity square $C_s^2$, the specific heat $C_v$, and the trace 
anomaly also show similar behavior as those in the case of second order phase transition. The only
difference is that the width of the cusp becomes very narrow. The downward cusp in the pressure 
density over energy density  $p_T/\epsilon_T$ and the sound velocity square $C_s^2$ becomes
a dip at $T_c$, and the upward cusp of the specific heat $C_v$ and the trace anomaly 
develops into a delta function at $T_c$. 
When the coupling strength increases, the value of  the pressure 
density over energy density $p_T/\epsilon_T$ and the sound velocity square $C_s^2$ 
at $T_c$ become smaller and smaller, while the value of the specific heat 
$C_v$ and the trace anomaly becomes bigger and bigger.
  
 The bulk viscosity over entropy density $\zeta/s$ decreases with $T$
at low temperature region, then sharply rises up at the critical temperature $T_c$,
and decreases further in the temperature $T>T_c$. The value of $\zeta/s$ 
becomes larger and larger at the critical temperature with the increase of
coupling strength.

\subsection{$O(4)$ model with spontaneous symmetry breaking  in the vacuum}

\begin{figure}[thbp]
\epsfxsize=7.5 cm \epsfysize=6.5cm \epsfbox{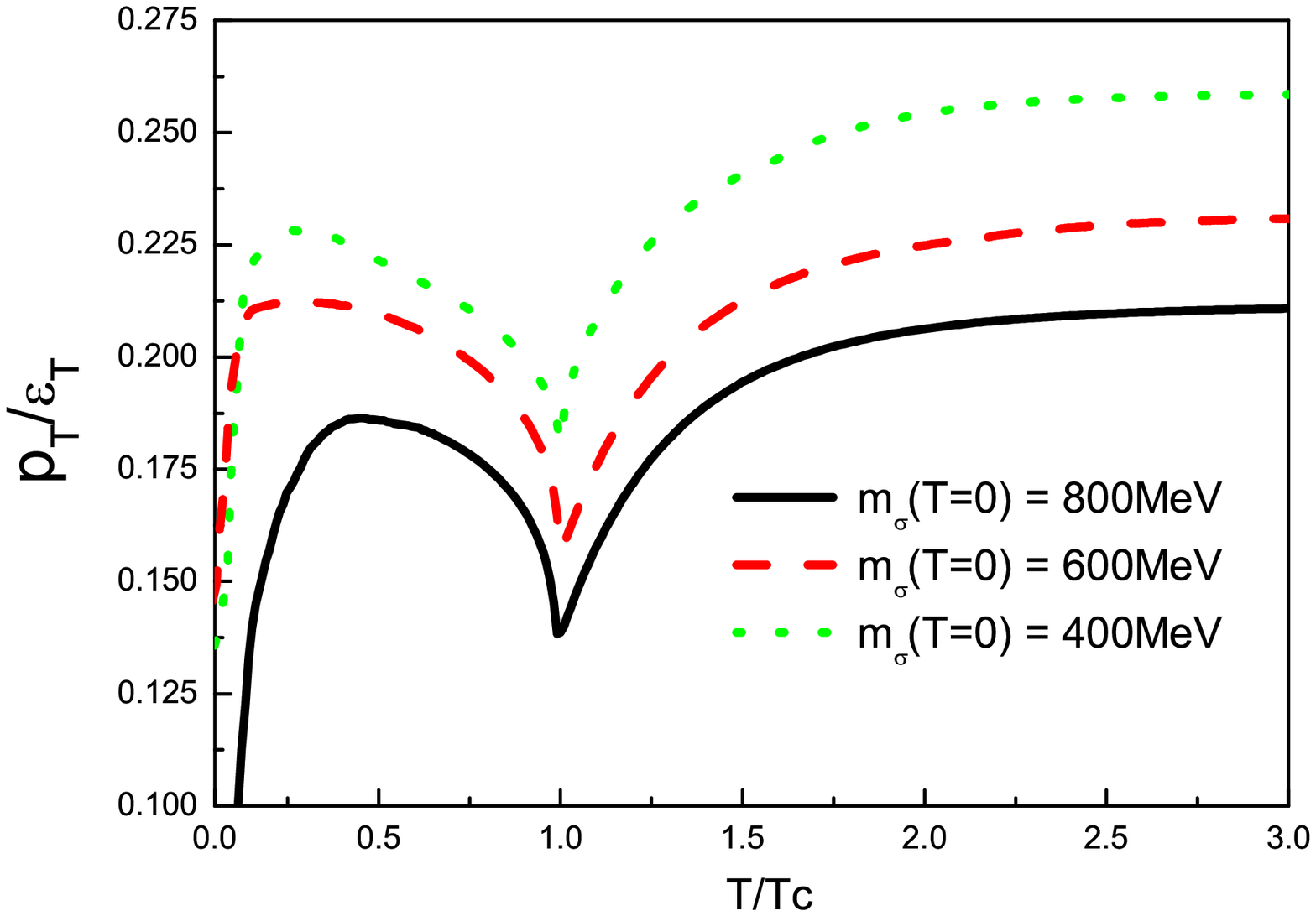}\hspace*{0.1cm}
\epsfxsize=7.5 cm \epsfysize=6.5cm \epsfbox{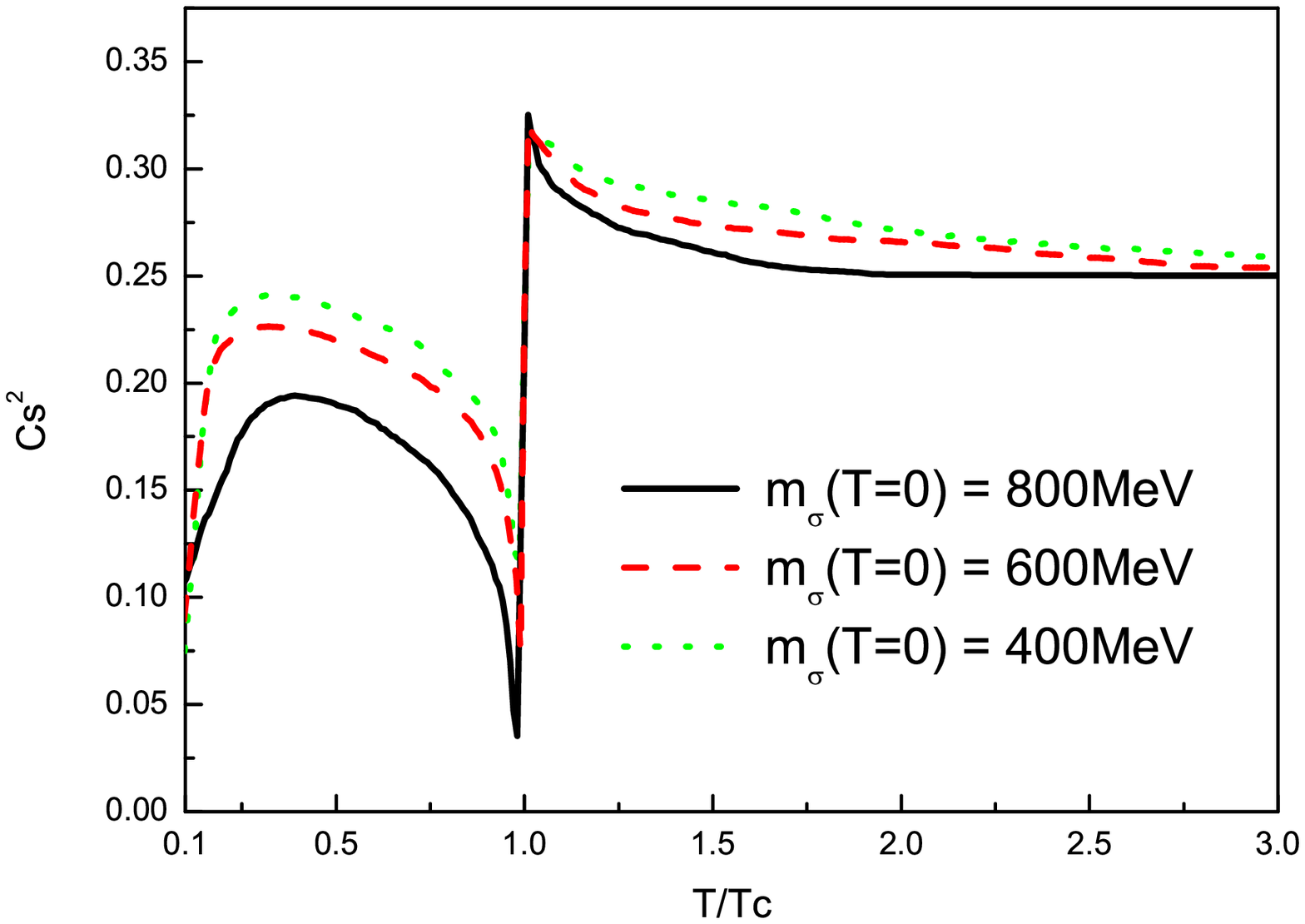}
\vskip -0.05cm \hskip 0.15 cm \textbf{( a ) } \hskip 6.5 cm \textbf{( b )} \\
\epsfxsize=7.5 cm \epsfysize=6.5cm \epsfbox{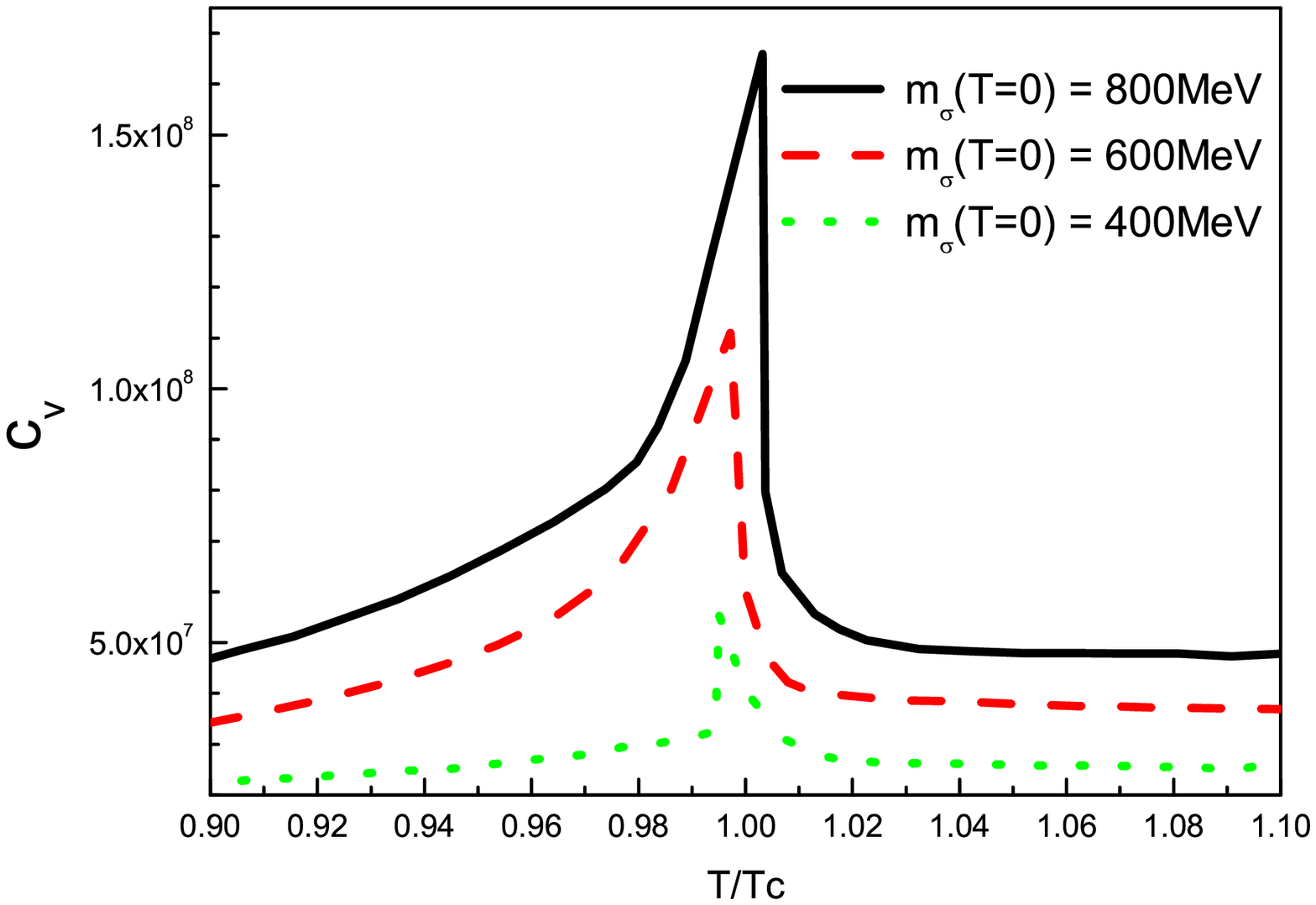} \hspace*{0.1cm}
\epsfxsize=7.5 cm \epsfysize=6.5cm \epsfbox{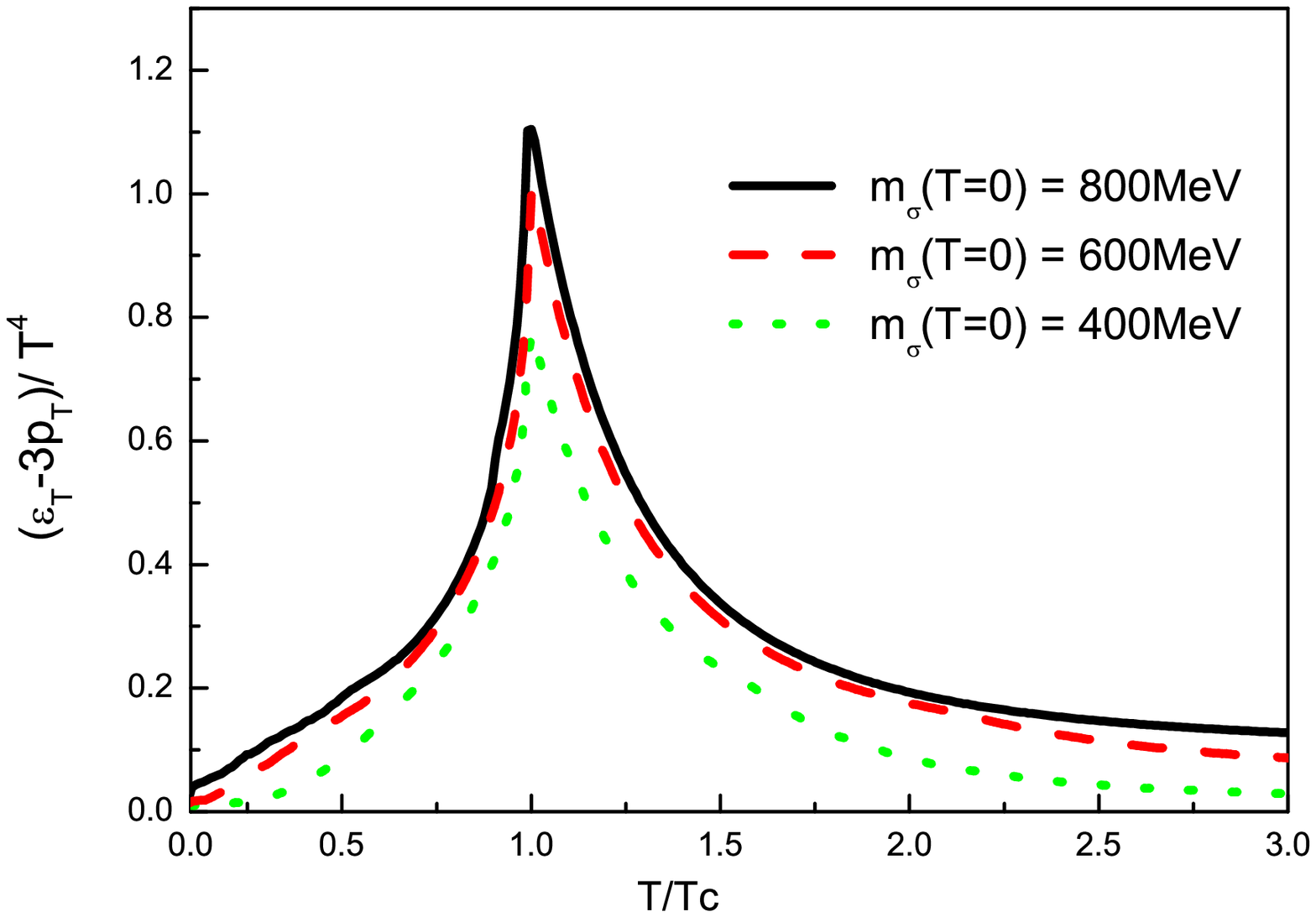} 
\vskip -0.05cm \hskip 0.15 cm \textbf{( c ) } \hskip 6.5 cm \textbf{( d )} \\
\epsfxsize=7.5 cm \epsfysize=6.5cm \epsfbox {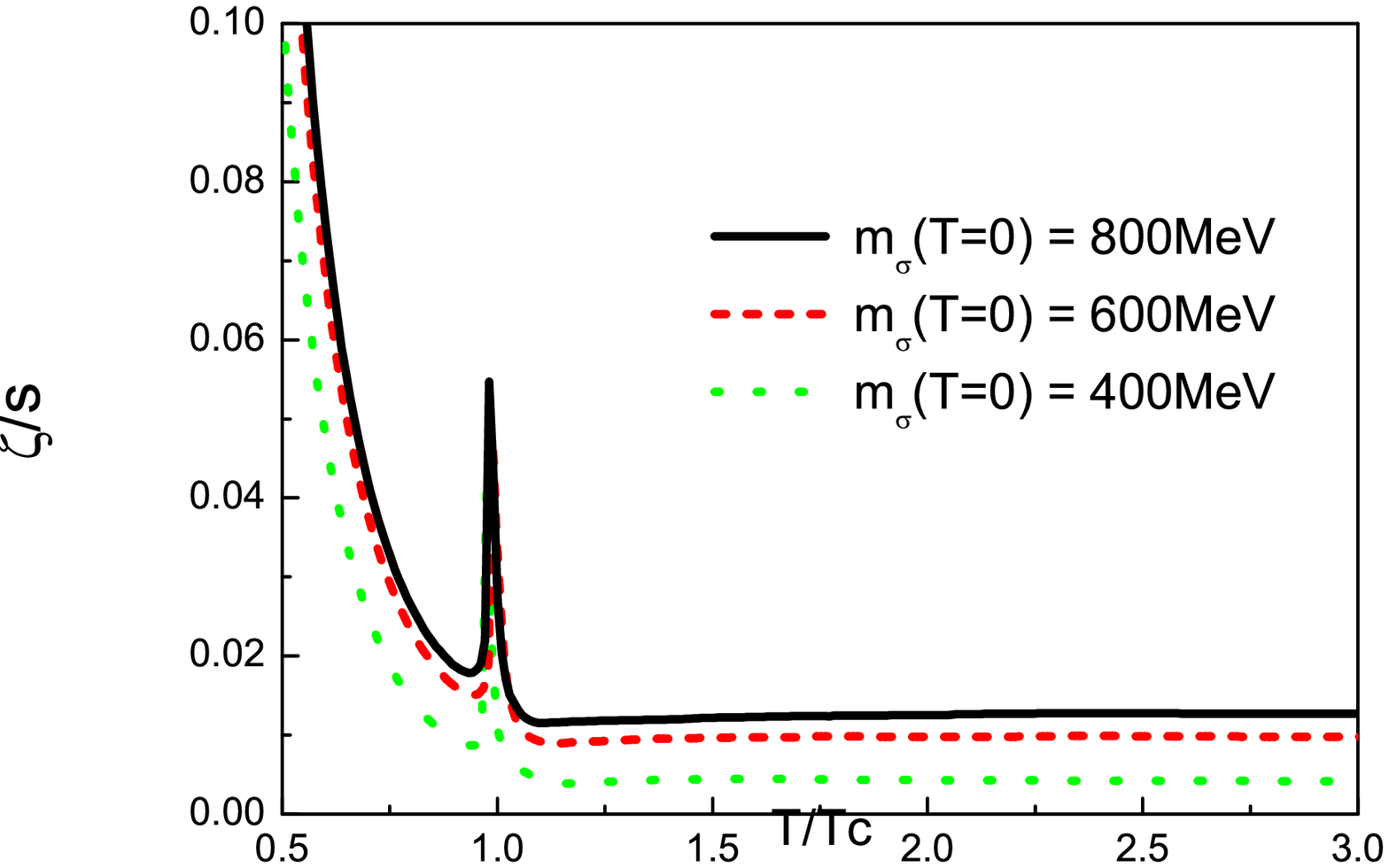} 
\vskip -0.05cm \hskip 0.15 cm \textbf{( e ) } 
\caption{The ratio of pressure density over energy density $p_T/\epsilon_T$, the trace anomaly 
$(\epsilon_T-3 p_T)/T^4$, the sound velocity square $C_s^2$, the specific heat 
$C_v$ and the bulk viscosity over entropy density ratio $\zeta/s$ as functions of the 
temperature $T$ with second order phase transition in the $O(4)$ model.}
\label{2nd_N=4}
\end{figure}

In Fig. \ref{2nd_N=4} $a-e$, we show the ratio of pressure density over energy density 
$p_T/\epsilon_T$, the sound velocity square $C_s^2$, the interaction measure 
$(\epsilon_T-3 p_T)/T^4$,  the specific heat $C_v$ and the bulk viscosity
over entropy density $\zeta/s$ as functions of the temperature $T$ for the $O(4)$ model 
with a second order phase transition in the chiral limit $H=0$. The parameters used for 
calculation are taken from \cite{Roder}: $1)\, a=-(282.84 {\rm MeV})^2, \, b=39.53$,
$2)\, a=-(424.264 {\rm MeV})^2, \, b=88.88$, and 
$3)\, a=-(565.685 {\rm MeV})^2, \, b=158.02$
which produce the vacuum pion mass $m_{\pi}=0$,  vacuum pion decay constant 
$f_{\pi}=90\, {\rm MeV}$, and vacuum sigma meson mass as $m_\sigma=400 {\rm MeV}$
$m_\sigma=600 {\rm MeV}$ and  $m_\sigma=800 {\rm MeV}$, respectively. 
The stronger coupling strength of $b$ corresponds to the larger sigma mass in the vacuum.
The critical temperature for the chiral symmetry restoration in these three cases are
$T_c=176 {\rm MeV}$. 

It is found that all the thermodynamic properties and bulk viscosity show similar behaviors
as those in the $Z(2)$ model with second order phase transition at strong coupling.

At high temperature region $T>T_c$, the behavior of the
the pressure density over energy density $p_T/\epsilon_T$,  the specific heat $C_v$, 
and the trace anomaly show similar behavior
as those in the symmetric case. $p_T/\epsilon_T$ increases with temperature 
and saturate at a value smaller than $1/3$, the stronger the coupling strength is, the smaller 
saturation value $p_T/\epsilon_T$ will be. The specific heat $C_v$ increases with temperature. 
The trace anomaly decreases with temperature and goes to a larger value for stronger coupling 
strength $b$. However, it is found that the sound velocity square $C_s^2$ shows different
behavior comparing with that in the $Z(2)$ case, $C_s^2$ decreases with the temperature
in the region $T>T_c$ in the $O(4)$ model and saturates at a value smaller than $1/3$.  

At low temperature region $T<T_c$, both the pressure density over energy density 
$p_T/\epsilon_T$ and the sound velocity square $C_s^2$ show a bump. 
There is no low-T peak showing up in the trace anomaly, because all these three
sets of parameters correspond to strong coupling.  

Near phase transition region $T\simeq T_c$, both the pressure density over energy density 
$p_T/\epsilon_T$ and the sound velocity square $C_s^2$ show a downward-cusp at $T_c$, 
the specific heat $C_v$ and the trace anomaly show an upward-cusp at $T_c$. 
When the coupling strength increases, the corresponding values of the pressure 
density over energy density $p_T/\epsilon_T$ and the sound velocity square $C_s^2$ 
at $T_c$ become smaller and smaller, while the critical values of the specific heat 
$C_v$ and the trace anomaly become larger and larger.
  
The bulk viscosity over entropy density $\zeta/s$ decreases with $T$
at low temperature region, then rises up at the critical temperature $T_c$.
The critical value of $\zeta/s$ at $T_c$ in the $O(4)$ model which simulates
the real QCD chiral phase transition is around $0.02-0.06$, which is rather
small comparing with the lattice QCD results.

\subsection{$O(4)$ model with explicit symmetry breaking}

\begin{figure}[thbp]
\epsfxsize=7.5 cm \epsfysize=6.5cm \epsfbox{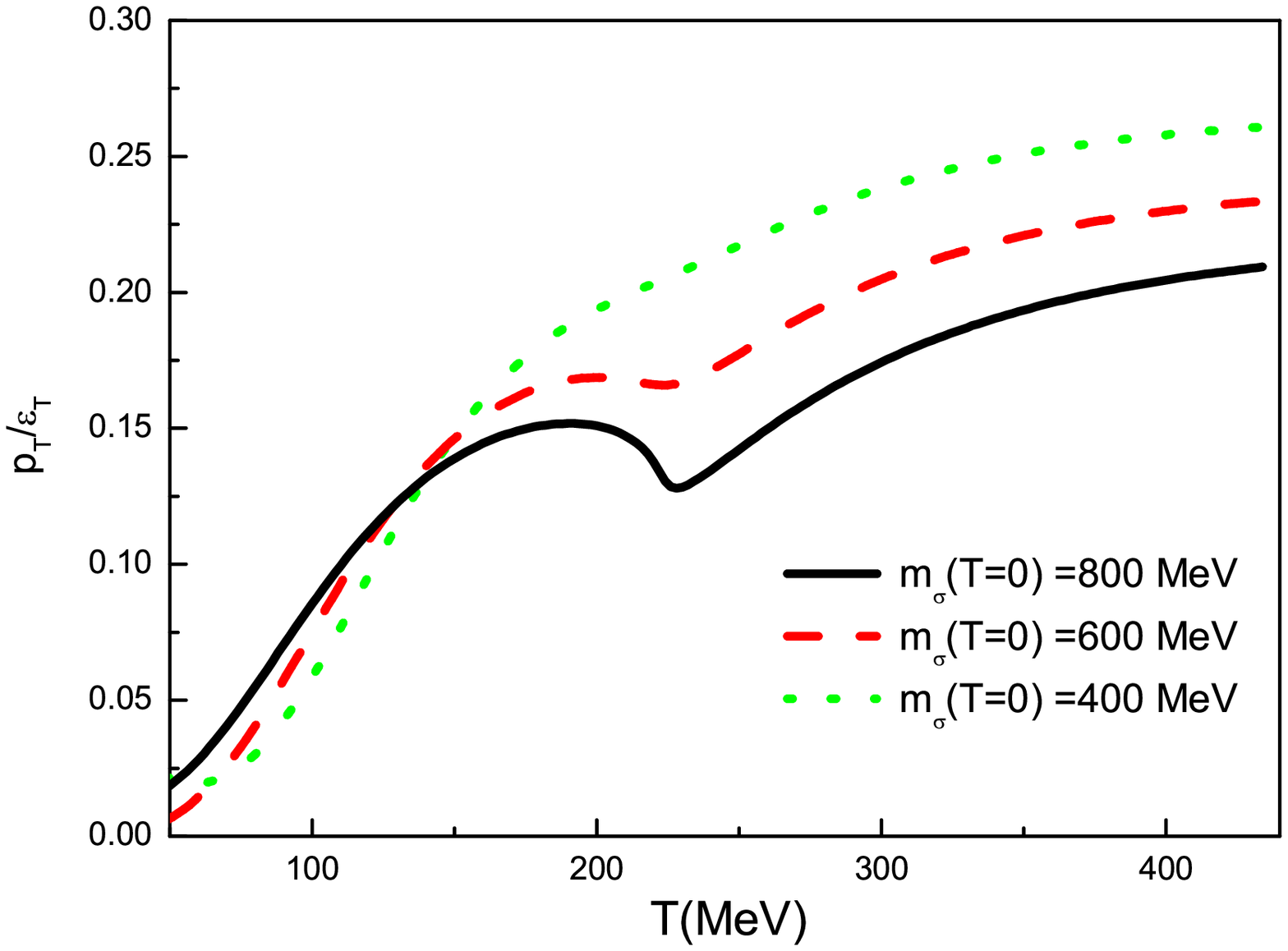}\hspace*{0.1cm}
\epsfxsize=7.5 cm \epsfysize=6.5cm \epsfbox{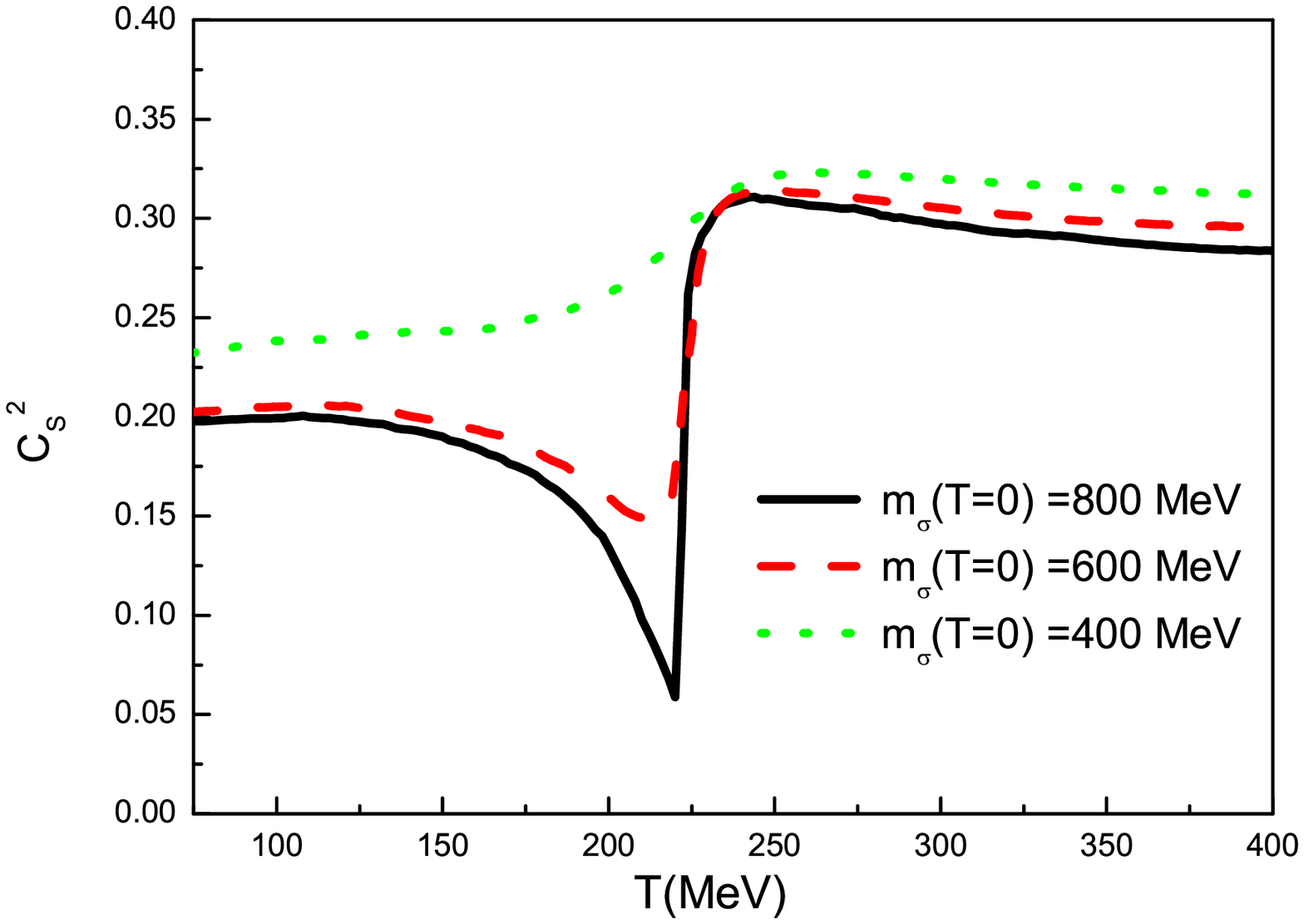}
\vskip -0.05cm \hskip 0.15 cm \textbf{( a ) } \hskip 6.5 cm \textbf{( b )} \\
\epsfxsize=7.5 cm \epsfysize=6.5cm \epsfbox{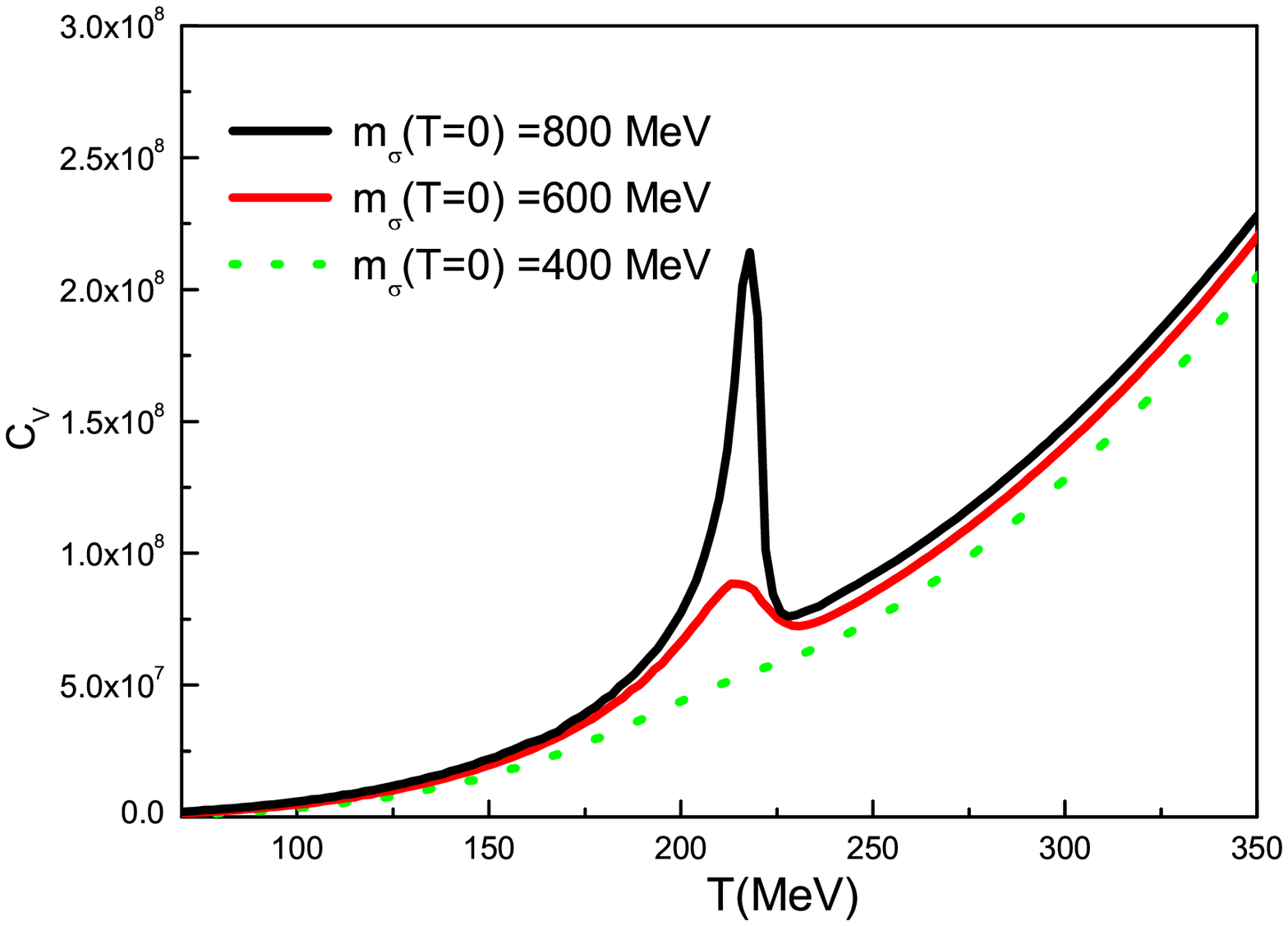}\hspace*{0.1cm}
\epsfxsize=7.5 cm \epsfysize=6.5cm \epsfbox{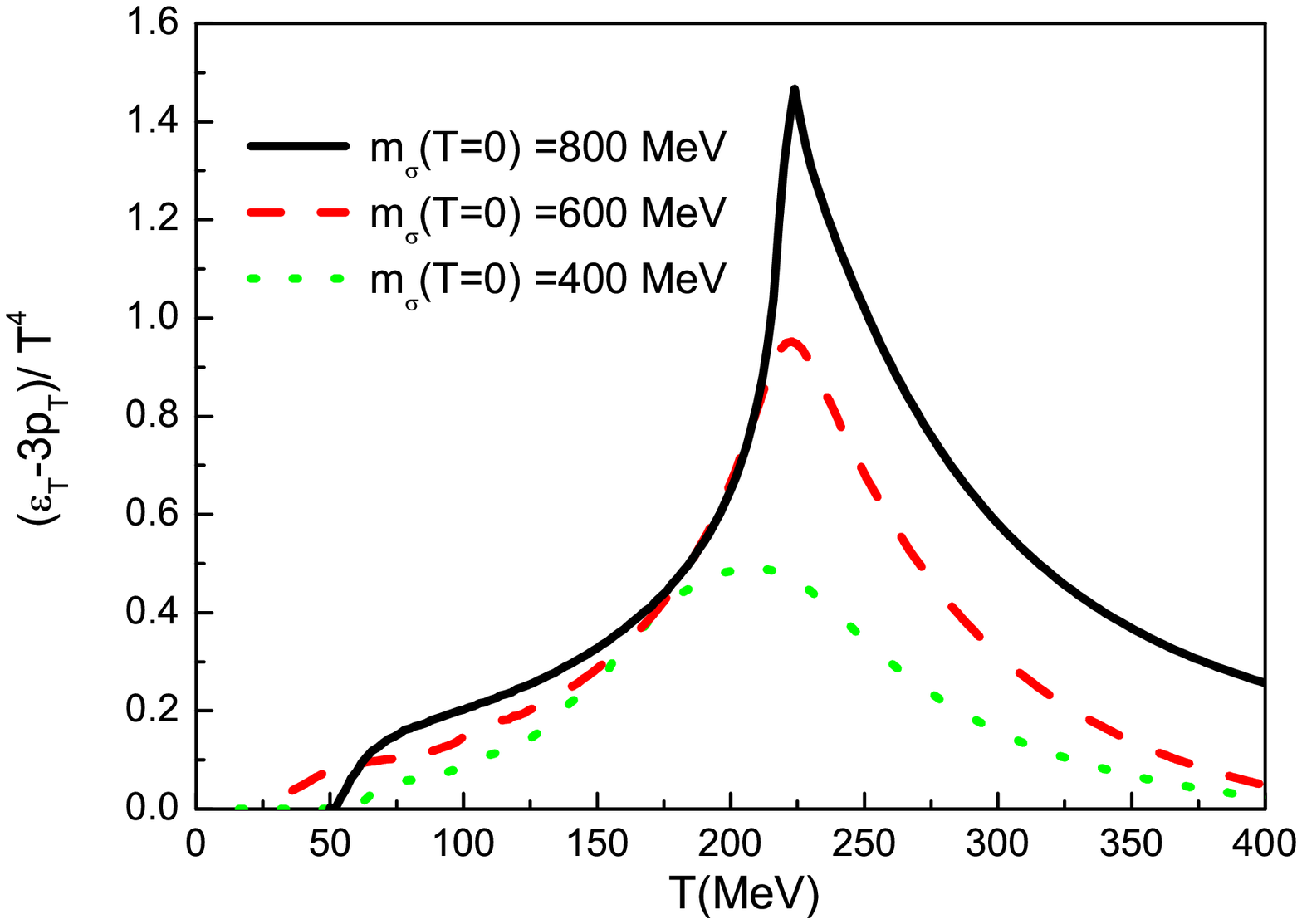}
\vskip -0.05cm \hskip 0.15 cm \textbf{( c ) } \hskip 6.5 cm \textbf{( d )} \\
\epsfxsize=7.5 cm \epsfysize=6.5cm \epsfbox {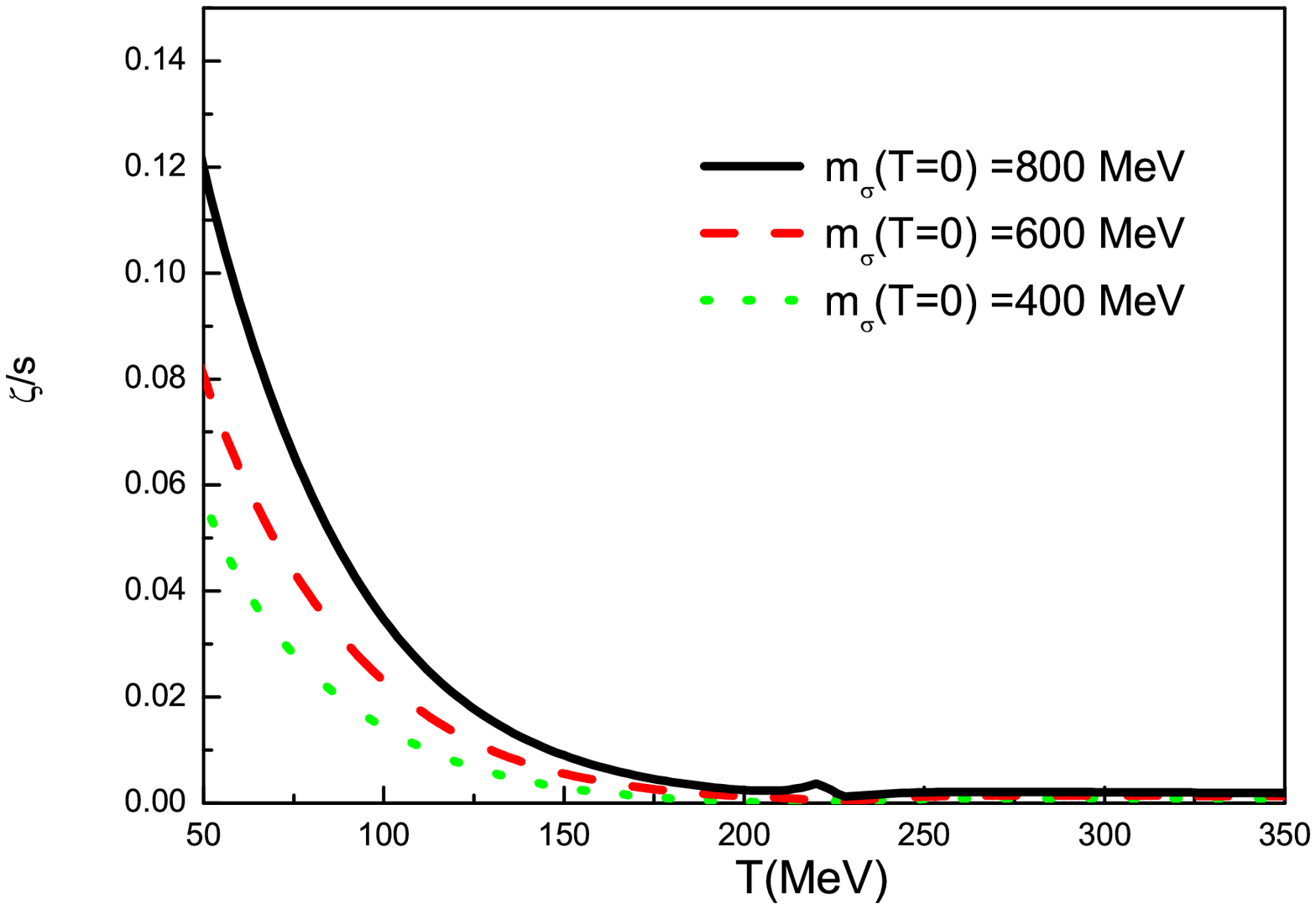} 
\vskip -0.05cm \hskip 0.15 cm \textbf{( e ) } 
\caption{The ratio of pressure density over energy density $p_T/\epsilon_T$, the 
trace anomaly $(\epsilon_T-3 p_T)/T^4$, the sound velocity square $C_s^2$, 
the specific heat $C_v$ and the bulk viscosity over entropy density ratio $\zeta/s$
as functions of the temperature $T$ with explicit symmetry breaking in the $O(4)$ model.
}
\label{Cross_N=4}
\end{figure}

In Fig. \ref{Cross_N=4} $a-e$, we show the ratio of pressure density over energy density 
$p_T/\epsilon_T$, the square of sound velocity $C_s^2$, the trace anomaly $(\epsilon_T-3 p_T)/T^4$,
the specific heat $C_v$ and the bulk viscosity over entropy density $\zeta/s$ as functions 
of the temperature $T$ for the $O(4)$ model in the case of crossover when chiral symmetry
is explicitly broken by a finite value $H=(121.6 {\rm MeV})^3$.  
The parameters used for calculation are taken from \cite{Roder}:
$1)\,a=-(225.41\, {\rm MeV})^2, \, b=32.12$,
$2)\,a=-(388.34 \,{\rm MeV})^2, \, b=76.17$,
and $3)\, a=-(539.27 {\rm MeV})^2, \, b=145.02$,
which produce the
vacuum pion mass $m_{\pi}=139.5 \,{\rm MeV}$,  the vacuum pion decay constant 
$f_{\pi}=92.4\, {\rm MeV}$, and vacuum sigma meson mass as $m_\sigma=400 {\rm MeV}$
$m_\sigma=600 {\rm MeV}$ and  $m_\sigma=800 {\rm MeV}$, respectively. 

At both low temperature region $T<T_c$ and high temperature region $T>T_c$, the 
behavior of all the thermodynamic quantities and bulk viscosity over entropy density
ratio show similar behavior as the case of second order phase transition in the $O(4)$ model.  
However, near critical temperature
region $T\simeq T_c$, it is observed all the cusp behaviors are washed out, e.g. 
the downward cusp in the pressure density over 
energy density $p_T/\epsilon_T$ and the sound velocity square $C_s^2$
develops into a shallow valley, and the upward cusp in the specific heat and
the trace anomaly develops into a smooth peak. 
The sharp rise of the bulk viscosity over entropy density $\zeta/s$ in the case $H=0$
vanishes in the chiral symmetry explicitly breaking case and there is no obvious 
change of $\zeta/s$ near the critical temperature.  
Together with the observation in the $Z(2)$ model, we can predict that at RHIC, 
where there is no real phase transition and 
the system experiences a crossover, the bulk viscosity over entropy 
density is small, and it will not affect too much on hadronization.

\section{Discussions and summary}
\label{sec-Summary}

\subsection{Comparing with other results in AdS/CFT, lattice QCD and effective QCD models}

We have investigated the thermodynamic properties and bulk viscosity in the $Z(2)$ and
$O(4)$ models in the Hartree approximation of CJT formalism. We now compare our results
with the results in AdS/CFT, lattice QCD and effective QCD models like the 
Polyakov-loop Nambu--Jona-Lasinio (PNJL) model.

The conformal limit has attracted much
attention in recent years, since people are trying to understand strongly
interacting quark-gluon plasma by using AdS/CFT techniques.
In conformal field theories including free field theory, the pressure density over
energy density and the sound velocity square is always $1/3$, i.e.
$p_T/\epsilon_T=c_s^2=1/3$, and the trace anomaly and the bulk viscosity is 
always zero, i.e. $\Delta=\zeta=0$.  Lattice results show that at asymptotically 
high temperature, the hot quark-gluon system is close to a conformal and free
ideal gas. Our results of $Z(2)$ and $O(4)$ models in the weak coupling also
reach the conformal limit at high temperature.

However, lattice results show that near deconfinement phase transition,
the hot quark-gluon system deviates far away from conformality.
Both $p_T/\epsilon_T$ and $c_s^2$ show a minimum  around $0.07$, which
is much smaller than $1/3$. For the $SU(3)$ pure gluon system\cite{LAT-EOS-G},
the peak value of the trace anomaly $\Delta_{LAT}^{G}$ reads $3\sim 4$ at
$T_{max}$ and the corresponding "interaction measure" is
$\Delta_{LAT}^{G}/d_G=0.2\sim 0.25$, with the gluon degeneracy factor $d_G=16$.
(Note that here $T_{max}\simeq 1.1 T_c$ is the temperature corresponding to the sharp peak
of $\Delta$.) For the two-flavor case \cite{LAT-EOS-Nf2}, the lattice result of the
peak value of the trace anomaly $\Delta_{LAT}^{Nf=2}$ reads $8\sim 11$,
the corresponding interaction measure at $T_{max}$ is given as
$\Delta_{LAT}^{Nf=2}/(d_G+d_Q)=0.28 \sim 0.4$, with quark degeneracy factor $d_Q=12$.
There have been some efforts trying to understand
trace anomaly in gluodynamics near and above $T_c$ in terms of dimension two gluon
condensate and an effective "fuzzy" bag model \cite{trace-anomaly}.

Our results in $Z(2)$ and $O(4)$ models show that at critical temperature $T_c$,
the trace anomaly $\Delta$, the specific heat $C_v$ as well as bulk viscosity to entropy
density ratio $\zeta/s$ show upward cusp at $T_c$, and their peak values increase with
the increase of coupling strength. The ratio of pressure density over energy density
$p_T/\epsilon_T$ and the square of the sound velocity $C_s^2$ show downward cusp at $T_c$,
which are similar to the behavior of $\eta/s$ found in Ref. \cite{etas-scalar}, and
the cusp values decrease with the increase of coupling strength.
These cusp behaviors at phase transition resemble lattice QCD results.

To our surprise, we find that when $b=30$, the strongly coulped real scalar system
can reproduce all thermodynamic and transport properties of hot quark-gluon system
near $T_c$. $p_T/\epsilon_T$ at ${T_c}$ is close to the lattice QCD result $0.07$,
$\Delta/d=0.48$ ($d=1$ for scalar system) at $T_c$ is close to the lattice result of the
peak value $\Delta_{LAT}^{Nf=2}/(d_G+d_Q)\simeq0.4$ at $T_{max}$. The bulk viscosity
to entropy density ratio $\zeta/s$ at $T_c$ is around $0.5\sim 2.0$, which agrees well with
the lattice result in Ref. \cite{LAT-xis-Meyer}. (Note, here $\zeta/s=0.5, 2$ correspond to
$\omega_0=10 T, 2.5 T$, respectively.) More surprisingly, the shear viscosity over entropy
density ratio $\eta/s$ at $T_c$ is $0.146$, which also beautifully agrees with lattice
result $0.1\sim 0.2$ in Ref. \cite{LAT-etas}.

Table \ref{table-all} shows our results of equation of state and transport
properties in the real scalar field theory or $Z(2)$ model at $T_c$ for 
second order phase transition, and corresponding results in lattice QCD calculations 
\cite{LAT-EOS-G,LAT-EOS-Nf2,LAT-xis-KT,LAT-xis-Meyer},
the PNJL model \cite{EOS-PNJL-Weise,EOS-PNJL-Ray},
and black hole duals \cite{Gubser-EOS}. In black hole solutions, it is found that
$\zeta/\eta\simeq 2 (1/3-C_s^2)$. In the real scalar model near phase transition, 
there is no any universal relationship between $\zeta$ and $\eta$.

\begin{table}[th]\vspace*{-0.0cm}
\begin{center}%
\begin{tabular}
[c]{|c|c|c|c|c|c|}\hline
  & $\eta/s$ & $\zeta/s$ & $\Delta/d$ & $ C_s^2$ & $p_T/\epsilon_T$ \\\hline
$b=30$  & $0.146 $ & $0.5\sim 2.0$ & $0.48$ & $0.03$ & $0.07$ \\\hline
${\rm LAT}_{G}$ \cite{LAT-EOS-G, LAT-xis-KT}& $0.1\sim 0.2$ & $0.5\sim 2.0$ & $0.25$ & $-$ & $0.07$ \\ \hline
${\rm LAT}_{Nf=2}$ \cite{LAT-EOS-Nf2, LAT-xis-KT} &  $-$ & $0.25\sim 1.0$ & $0.4$ & $0.05$ & $0.07$ \\ \hline
${\rm PNJL}$ \cite{EOS-PNJL-Weise,EOS-PNJL-Ray} &  $ - $ & $-$ & $0.21$ & $0.08$ & $0.075$ \\ \hline
${\rm AdS/CFT}$ \cite{bound} &  $ 1/4\pi $ & $0$ & $0$ & $1/3$ & $1/3$ \\ \hline
${\rm Type I BH}$ \cite{Gubser-EOS} &  $ 1/4\pi $ & $0.06$ & $-$ & $0.05$ & $-$ \\ \hline
${\rm Type II BH}$ \cite{Gubser-EOS} &  $ 1/4\pi $ & $0.08$ & $-$ & $\simeq 0$ & $-$ \\ \hline
\end{tabular}
\end{center}
\vspace*{-0.1cm}
\caption{Thermodynamic and transport properties at $T/T_c=1$ in $Z(2)$ at 
strong coupling $b=30$, in lattice QCD
\cite{LAT-EOS-G,LAT-EOS-Nf2,LAT-xis-KT,LAT-xis-Meyer}, PNJL model
\cite{EOS-PNJL-Weise,EOS-PNJL-Ray}, and black hole dules \cite{Gubser-EOS}. The
degeneracy factor $d=1,16,28$ for $Z(2)$ model, pure gluon system, and 2-flavor
quark-gluon system, respectively. Lattice results are taken at $T_{max}$.  }%
\label{table-all}%
\vspace*{-0.15cm}
\end{table}

\subsection{Low-T peak of the trace anomaly}

In the $Z(2)$ model, it is found that the trace anomay $(\epsilon_T-3 p_T)/T^4$ 
shows a peak at low temperature in the case of weak coupling, and this low-T 
peak disappears in the strong coupling case. 

The peak of the trace anomaly $(\epsilon_T-3 p_T)/T^4$ at low temperature, which is not 
related to the phase transition,  was also observed in Ref. \cite{Bulk-Nicola} in the Chiral 
Perturbation Theory for the pion gas. In Ref. \cite{Bulk-Nicola}, the low-T peak of the trace 
anomaly was interpreted as the explicit conformal breaking, whose contribution comes from 
massive pions. However, for the real scalar system, there are no massive pions, it is not clear 
for us what is the reason inducing the conformal symmetry breaking at low temperature.

We don't observe the correlation between the low-T peak of the trace anomaly and the
bulk viscosity. It might be due to the method we have used to calculate the bulk viscosity.  It is
worthy of checking whether the correlation really exists by using Kubo formula to calculate
the bulk viscosity. 

\subsection{$\zeta/s$ at RHIC} 
\label{zetas-RHIC}

Recent lattice QCD results show that the bulk viscosity over
entropy density ratio $\zeta/s$ rises dramatically up to the order of $1.0$ near the
critical temperature $T_c$ \cite{LAT-xis-KT,LAT-xis-Meyer}. 
The sharp rise of the bulk viscosity will lead to the breakdown of the
hydrodynamic approximation around the critical temperature, and will
affect the hadronization and freeze-out processes of QGP created at heavy ion collisions.
The authors of Ref. \cite{bulk-Mishustin} pointed out the possibility that a sharp
rise of bulk viscosity near phase transition induces an instability in the hydrodynamic
flow of the plasma, and this mode will blow up and tear the system into droplets.
Another scenario is pointed out in Ref. \cite{LAT-xis-KT,bulk-review-Kharzeev} that the large
bulk viscosity near phase transition might induce ``soft statistical hadronization". 

However, if the strongly coupled QGP created at RHIC experiences a crossover and
not a real phase transition, from our results in the simple $Z(2)$ and $O(4)$ models,
the sharp rise of the bulk viscosity over entropy density will be washed out. In this case, 
one does not need to worry about the breakdown of the hydrodynamic approximation 
around the critical temperature. 

\subsection{Using $\zeta/s$ to locate the CEP}

For real QCD with two quarks of small mass, it is expected that there exists a 
critical end point (CEP) in the $T-\mu$ QCD phase diagram. 
At small baryon chemical potential $\mu$, the chiral phase transition is  a 
smooth crossover at finite temperature.  
At finite baryon chemical potential,  the chiral phase transition 
is of first order. The precise location of the CEP is still unknown. In the future plan, 
RHIC is going to lower the energy and trying to locate the CEP. Recently, the authors of 
Ref. \cite{etas-CEP} suggested using the shear viscosity over entropy 
density ratio $\eta/s$ to locate the CEP.

From Refs.\cite{etas-scalar, Csernai:2006zz}, we know that $\eta/s$ shows a shallow
valley in the case of crossover and a jump at $T_c$ for first-order phase transition. 
But it is hard to distinguish whether the system experiences a 
crossover or first-order phase transition just from the value of $\eta/s$ extracted from
the elliptic flow $v_2$.

From our results in $Z(2)$ and $O(4)$ models, it is found that the ratio of $\zeta/s$ shows
a very sharp peak at $T_c$, and there is no obvious change of $\zeta/s$ for crossover. 
As pointed out in Ref. \cite{bulk-Mishustin} that a sharp
rise of bulk viscosity near phase transition induces an instability in the hydrodynamic
flow of the plasma, and this mode will blow up and tear the system into droplets. 
Therefore, one can distinguish whether the system experiences a first order phase transition
or a crossover from observables  at RHIC experiments.

Therefore, $\zeta/s$ is a better quantity than $\eta/s$ to locate the CEP. (It is noticed that
because the QGP created at RHIC is a finite system, we don't discuss the singularity of 
$\eta/s, \zeta/s$ at CEP. )

\subsection{Limitations of our results}

At the end, we hope to point out some limitations of our results.

Firstly, the results of thermodynamic properties in this
paper are based on Hartree approximation in the CJT formalism. As we know that mean-field
approximation cannot describe critical phenomena very well. For
2nd-order phase transition in $Z(2)$ model, the specific heat $C_v$ should diverge
at the critical point, and behave as $t^{-\alpha}$ near the critical point,  with $t=(T-T_c)/T_c$
and $\alpha=0.11$. However, in Hartree approximation of CJT formalism,
though we observe the weak divergence of $C_v$ at $T_c$ in the case of strong coupling, we
can only see a weak upward cusp of $C_v$ at $T_c$ in the case of weak coupling.
As pointed out in the introduction, the QGP created at RHIC is a finite system, the 
singularity of $\eta/s, \zeta/s$ at critical point will not show up in the observables. If we are only
interested in the qualitative properties near phase transition, the Hartree 
approximation can give the dominant contributions.
 
Secondly, the results of bulk viscosity in this paper are based on Eq. (\ref{ze}).
The limitation of Eq. (\ref{ze}) has been analyzed in Refs. \cite{Moore} and
\cite{correlation-Karsch}. From Eq. (\ref{ze}), we see that the bulk viscosity is dominated
by $C_v$ at $T_c$. If $C_v$ diverges at $T_c$, the bulk viscosity should also be divergent
at the critical point and behave as $t^{-\alpha}$. However, the detailed analysis in the Ising
model in Ref. \cite{Onuki} shows a very different divergent behavior
$\zeta \sim t^{-z\nu +\alpha}$, with $z\simeq 3 $ the dynamic
critical exponent and $\nu\simeq 0.630$ the critical exponent in the Ising system.

Thirdly, we should keep in mind that our results at strong coupling are from effective 
theory. From renormalization analysis, the scalar theory will hit a Landau pole when 
$\beta_b > b$ with the $\beta$ function $\beta_b = 9 b^2/16\pi^2$. The results for large 
$b$ in the scalar theory are not guaranteed to be valid in the CJT formalism.

More careful study on thermodynamic properties in this model beyond mean-field approximation
is needed, and  the bulk viscosity with full spectral function of the pressure-pressure
correlator is in progress.

\subsection{Summary}

In summary, in the Hartree approximation of CJT formalism, we have investigated the thermodynamic
properties and transport properties of the $Z(2)$ model and $O(4)$ model, and compared
these properties in the cases of first and second order phase transitions, in the case
of crossover and the case of symmetric phase.

We have seen that at phase transition, the system either in weak coupling or strong coupling shows
some common properties. 1)  The pressure density over energy density ratio $p_T/\epsilon_T$, 
the square of the speed of sound $C_s^2$ as well as
$\eta/s$ exhibit downward cusp behavior at $T_c$. 2) The trace anomaly $\Delta$, the specific heat
$C_v$ as well as $\zeta/s$ show upward cusp behavior at $T_c$. The cusp behavior is related to the 
biggest change rate of entropy density at $T_c$ \cite{Asakawa-Hatsuda}. 3) The cusp behavior in 
the first order phase transition is sharper and narrower than that in the second order phase transition. 
4) In the case of crossover, the cusp behavior is washed out. 

Therefore,
if the strongly coupled QGP created at RHIC experiences a crossover and
not a real phase transition, one does not need to worry about the breakdown 
of the hydrodynamic approximation around the critical temperature.
Because the behavior of $\zeta/s$ is so different in the case of first order phase transition
and the crossover, we also suggest that $\zeta/s$ is a better quantity than $\eta/s$ to 
locate the CEP.

\section*{Acknowledgments} This work is supported by CAS program
"Outstanding young scientists abroad brought-in", CAS key project KJCX3-SYW-N2,
NSFC10735040, NSFC10875134 and NSFC10675077. 


\end{document}